\documentclass[12pt]{JHEP3}

\usepackage{graphicx}
\usepackage{cite}
 \usepackage{ae}
 \usepackage{aecompl}
\usepackage{amsmath,epsfig}
\usepackage{amssymb,amsfonts}
\usepackage{latexsym}
\usepackage{epsfig}
\usepackage{subfigure}

\relax
\renewcommand{\theequation}{\arabic{section}.\arabic{equation}}

\def\be{\begin{equation}}
\def\ee{\end{equation}}

\newcommand{\bear}{\begin{eqnarray}}
\newcommand{\bea}{\begin{eqnarray}}
\newcommand{\eear}{\end{eqnarray}}
\newcommand{\eea}{\end{eqnarray}}

\newbox\pippobox

\def\II{\relax{\rm I\kern-.18em I}}

\def\cO{{\cal O}}

\def\l{\lambda}
\def\m{\mu}
\def\n{\nu}

\def\a{\alpha}

\def\lab{\label}
\def\o{\omega}

\def\gcs{\Gamma_{\textrm{CS}}}
\def\ncs{N_{\textrm{CS}}}
\def\N{{\mathcal{N}}}

\def\k{\kappa}
\def\E{{\mathcal{E}}}

\title{The Chern-Simons Diffusion Rate in Improved Holographic QCD}

\author{U.~G\"ursoy$^{a,b}$, I.~Iatrakis$^c$, E.~Kiritsis$^{c,d}$, F.~Nitti$^d$, A.~O'Bannon$^e$
~\\
$^a$Institute for Theoretical Physics, Utrecht University\\ Leuvenlaan 4, 3584 CE Utrecht, The Netherlands
~\\
$^b$Instituut voor Theoretische Fysica, KU Leuven\\ Celestijnenlaan 200D, B-3001 Leuven, Belgium
~\\
$^c$Crete Center for Theoretical Physics, Department of Physics\\ University of Crete, PO Box 2208, 71003 Heraklion, Greece
~\\
$^d$APC, Universit\'e Paris 7, CNRS/IN2P3, CEA/IRFU, Obs. de Paris, Sorbonne Paris Cit\'e, B\^atiment Condorcet, F-75205, Paris Cedex 13, France (UMR du CNRS 7164)
~\\
$^e$Department of Applied Mathematics and Theoretical Physics \\
University of Cambridge, Cambridge CB3 0WA, United Kingdom
~\\
}

\preprint{CCTP-2012-20, DAMTP-2012-75 \\ ITP-UU-12/45, SPIN-12/42}

\abstract{In $(3+1)$-dimensional $SU(N_c)$ Yang-Mills (YM) theory, the Chern-Simons diffusion rate, $\Gamma_{CS}$, is determined by the zero-momentum, zero-frequency limit of the retarded two-point function of the CP-odd operator $\textrm{tr} \left[F \wedge F\right]$, with $F$ the YM field strength. The Chern-Simons diffusion rate is a crucial ingredient for many CP-odd phenomena, including the chiral magnetic effect in the quark-gluon plasma. We compute $\Gamma_{CS}$ in the high-temperature, deconfined phase of Improved Holographic QCD, a refined holographic model for large-$N_c$ YM theory. Our result for $\Gamma_{CS}/(sT)$, where $s$ is entropy density and $T$ is temperature, varies slowly at high $T$ and increases monotonically as $T$ approaches the transition temperature from above. We also study the retarded two-point function of $\textrm{tr}\left[F \wedge F\right]$ with non-zero frequency and momentum. Our results suggest that the CP-odd phenomena that may potentially occur in heavy ion collisions could be controlled by an excitation with energy on the order of the lightest axial glueball mass.
}

\begin{document}

\maketitle 

\section{Introduction}

(3+1)-dimensional $SU(N_c)$ Yang-Mills (YM) theory has an infinite number of degenerate classical vacua distinguished by a topological invariant, the Chern-Simons number, $N_{CS}$. Normalizing the YM kinetic term as $-\frac{1}{4g^2}\textrm{tr}[F_{\m\n}F^{\m\n}]$, $N_{CS}$ is
\be
\ncs \equiv \frac{1}{8\pi^2} \int d^3x \, \epsilon_{ijk} \, \textrm{tr} \left[ A_i \partial_j A_k - \frac{2ig}{3} A_iA_jA_k\right],
\ee
where $i,j,k=1,2,3$ and the trace is over gauge indices. A change in the Chern-Simons number is thus
\begin{subequations}
\be
\Delta \ncs = \int d^4 x \, q(x^{\mu}),
\ee
\be
\label{q}
q(x^{\mu})\equiv \frac{1}{16\pi^2}\textrm{tr} \left[F \wedge F\right] = \frac{1}{64\pi^2} \, \epsilon^{\mu\nu\rho\sigma} \textrm{tr} F_{\mu\nu} F_{\rho\sigma},
\ee
\end{subequations}
where $x^{\mu}=(t,\vec{x})$. In a state invariant under translations in space and time, the rate of change of $\ncs$ per unit volume $V$ per unit time $t$ is called the Chern-Simons diffusion rate, denoted $\gcs$,
\be
\label{gcsdef}
\gcs \equiv \frac{\langle \left(\Delta \ncs\right)^2\rangle}{V t} = \int d^4x \, \left \langle q(x^{\mu}) q(0) \right \rangle_{\textrm{W}},
\ee
where the subscript W denotes the Wightman function. In an equilibrium state with non-zero temperature $T$, let $G_R(\omega,\vec{k})$ denote the retarded Green's function of $q(x^{\mu})$ in Fourier space, with frequency $\omega$ and spatial momentum $\vec{k}$. In such states, eq.~\eqref{gcsdef} can be rewritten as
\be
\label{gcs0}
\Gamma_{CS}=-\lim_{\omega \to 0}\,\frac{2T}{\omega}\,\textrm{Im}\, G_R (\omega,\vec{k}=0).
\ee

Gauge field configurations for which $\int d^4x\,q(x^{\mu})$ is non-zero produce a non-zero $\Delta \ncs$. At zero temperature, such gauge field configurations, called instantons, represent quantum tunneling events between vacua. At both zero and non-zero $T$, the contribution of instantons to $\gcs$ is exponentially suppressed~\cite{Shuryak:1978yk,Gross:1980br}. When $T$ is non-zero, however, classical thermal fluctuations can also produce a non-zero $\Delta \ncs$, for example by exciting unstable gauge field configurations called sphalerons~\cite{Manton:1983nd,Klinkhamer:1984di} which generate non-zero $\Delta \ncs$ upon decay. Such classical thermal processes are not exponentially suppressed~\cite{Kuzmin:1985mm,Arnold:1987mh,Arnold:1987zg}: in YM perturbation theory $\gcs \propto \lambda_t^5 \log(\lambda_t)\, T^4$, where $\lambda_t \equiv g^2N_c$ is the 't Hooft coupling~\cite{Arnold:1996dy,Bodeker:1998hm,Bodeker:1999gx,Moore:2010jd}.

In YM coupled to fundamental-representation fermions, $\int d^4x\,q(x^{\mu})$ also contributes to chiral anomalies in global symmetries. In the electroweak theory, gauge field configurations with non-zero $\int d^4x\,q(x^{\mu})$ play a role in electroweak baryogenesis~\cite{Cohen:1993nk,Rubakov:1996vz}, while in Quantum Chromodynamics (QCD), for sufficiently high $T$ they may play a role in generating bubbles of net chirality (more left-handed than right-handed quarks, for example), in which parity, P, and charge conjugation times parity, CP, are broken~\cite{Kharzeev:2000na}.\footnote{Sometimes $\gcs$ is also called the ``sphaleron transition rate'' or, in the context of electroweak baryogenesis, the ``baryon number violation rate.''}

Such CP-odd domains in hot QCD may have observable consequences in heavy ion collisions at the Relativistic Heavy Ion Collider (RHIC) and Large Hadron Collider (LHC). These collisions produce a hot soup of QCD matter with $T$ on the order of two to four times the QCD crossover temperature. The resulting state behaves as a nearly-ideal fluid of strongly-interacting quarks and gluons, the quark-gluon plasma (QGP)~\cite{Gyulassy:2004zy,Shuryak:2008eq,Muller:2012zq}. A non-central collision may produce a QGP with non-zero angular momentum and hence a magnetic field, both pointing perpendicular to the reaction plane (spanned by the beam axis and impact parameter). In the presence of a magnetic field, a net chirality will produce an electric current parallel to the magnetic field, due to the axial anomaly. This is the Chiral Magnetic Effect (CME)~\cite{Kharzeev:2007jp,Fukushima:2008xe}. A detection of the CME in heavy ion collisions would thus be a detection of CP-odd processes in QCD.

One observable consequence of the CME in a heavy ion collision is charge separation: positive charges will move to one side of the reaction plane, negative charges to the other. We know from experiment that the strong interactions preserve P and CP, however, so any charge separation from CP-odd sources will, over many events, average to zero. An observable that could serve as a ``smoking gun'' for the CME is thus hard to find. For heavy ion collsions at RHIC and LHC the focus so far has been on three-particle correlations~\cite{Voloshin:2004vk,Abelev:2009ad,Abelev:2012pa}, which indeed indicate that charge separation occurs in heavy ion collisions. These correlations are sensitive to event-by-event charge separation from both CP-odd and CP-even processes, however, making a positive identification of a signal from the CME difficult~\cite{Bzdak:2012ia}. In short, to date the experimental evidence for the detection of the CME in heavy ion collisions at RHIC and LHC is inconclusive.

The experimental situation raises a number of urgent questions for theorists. Can we compute the size of the signal from the CME, relative to backgrounds? How will that signal depend on temperature, magnetic field, centrality, etc.? Clearly an auxiliary question is: how big is the rate of chirality production, which is $\propto \gcs$, in a heavy ion collision?

Unfortunately, $\gcs$ is difficult to calculate for the QGP, for the same reasons that the shear viscosity, $\eta$, is difficult to calculate. The quarks and gluons are strongly-interacting, so perturbation theory is \textit{a priori} unreliable. Calculating transport coefficients, such as $\gcs$ and $\eta$, from lattice QCD requires a problematic analytic continuation from Euclidean signature.\footnote{In fact, $\gcs$ may be \textit{more} difficult to calculate from lattice QCD than other transport coefficients. The operator $q(x^{\mu})$ obeys various constraints. For example, $\Delta N_{CS}=\int d^4x\,q(x^{\mu})$ must be an integer, the second Chern character. Defining a lattice version of the operator $q(x^{\mu})$ that obeys all of the constraints can be difficult, as discussed for example in refs.~\cite{Vicari:2008jw,Iqbal:2009xz,Moore:2010jd}.} Currently no reliable technique exists to compute $\gcs$ or $\eta$ for QCD at the temperatures reached in the QGP.

An alternative approach is holography~\cite{Maldacena:1997re,Gubser:1998bc,Witten:1998qj}, which equates certain strongly-coupled gauge theories in the large-$N_c$ limit with weakly-coupled theories of gravity in spacetimes of one higher spatial dimension. A deconfined thermal state of the gauge theory is dual to a black hole spacetime~\cite{Witten:1998zw}, and transport coefficients are relatively straightforward to calculate~\cite{Son:2002sd,Policastro:2002se,Iqbal:2008by}. Remarkably, the ratio of $\eta$ to entropy density, $s$, for \textit{any} theory dual to higher-dimensional Einstein gravity is $\eta/s = 1 / (4\pi)$~\cite{Kovtun:2004de}, which is close to the estimate for $\eta/s$ for the QGP extracted from data~\cite{Luzum:2008cw}. Such universality serves as encouragement for computing other transport coefficients, like $\gcs$, from holography.

Previous calculations of $\gcs$ in holography employed ``top-down'' models, \textit{i.e.}\ models descending from a known string theory or supergravity construction. The best-understood example is $\N=4$ supersymmetric YM (SYM) with large $N_c$ and large $\lambda_t$, dual to supergravity in an Anti-de Sitter (AdS) space, for which $\gcs \propto \lambda_t^2\,T^4$~\cite{Son:2002sd}. Other holographic calculations included the effects on $\gcs$ due to a magnetic field~\cite{Basar:2012gh} or confinement~\cite{Craps:2012hd}. To our knowledge, in all previous cases the holographic results for $\gcs$ were ultimately fixed by some underlying (perhaps ``hidden''~\cite{Craps:2012hd}) conformal symmetry.

In this paper we compute $\gcs$ in Improved Holographic QCD (IHQCD)~\cite{Gursoy:2007cb,Gursoy:2007er,Gursoy:2008bu,Gursoy:2008za,Kiritsis:2009hu,Gursoy:2009jd,Gursoy:2009kk,Gursoy:2010fj}, a holographic model of large-$N_c$ YM theory. The model is ``bottom-up," \textit{i.e.}\ does not descend from a known string theory or supergravity construction, but is tailored to model string theory systems very closely, unlike other bottom-up models. The bulk theory is Einstein-dilaton gravity, where the dilaton $\Phi$ is dual to $\textrm{tr} \, F_{\mu\nu}F^{\mu\nu}$. A non-trivial dilaton solution will describe non-trivial running of the YM coupling, hence the choice of dilaton potential is crucial. The simplest choice involves only two free parameters, which can be adjusted such that the model reproduces both the $T=0$ glueball spectrum and the thermodynamics of large-$N_c$ YM, including a first-order deconfinement transition at a critical temperature $T_c$. In particular, the model has no (hidden) conformal symmetry. We briefly review IHQCD in section~\ref{review}.

In IHQCD the operator $q(x^{\mu})$ is dual holographically to an axion field in the bulk~\cite{Gursoy:2007cb,Gursoy:2007er,Gursoy:2008za,Gursoy:2009jd,Gursoy:2010fj}. Defining for convenience a holographic 't Hooft coupling $\lambda \equiv e^{\Phi}$, the normalization of the axion's kinetic term includes a dilaton-dependent factor, $Z(\lambda)$. In principle, $Z(\lambda)$ could be fixed by matching to lattice results for the Euclidean correlator of $q(x^{\mu})$, as we explain in section~\ref{review}. We work instead with several simple choices for $Z(\lambda)$, in part to study the generic behavior of $\gcs$ in holographic models. Specifically, we consider a $Z(\lambda)$ with two free parameters, which we fix by demanding that the model match large-$N_c$ YM lattice results for the topological susceptibility and for axial glueball mass ratios to within one sigma.

In section~\ref{GammaCS} we compute $\gcs$ in the high-temperature, deconfined phase of IHQCD. Letting $s$ denote the entropy density and $\lambda_h$ the value of $\lambda$ at the black hole horizon, our result for $\gcs$ is of the form
\be
\label{gammaCS1}
\Gamma_{CS}= \frac{1}{N_c^2}{sT \over 2 \pi}\, Z(\lambda_h).
\ee
Figs.~\ref{fig3},~\ref{fig3b}, and~\ref{sigmafig} show our numerical results for $\gcs/(sT/N_c^2)=Z(\lambda_h)/(2\pi)$. For our choices of $Z(\lambda)$, the value of $Z(\lambda_h)$ is bounded from below as a function of $T$ by its value in the $T\to\infty$ limit, and increases monotonically as $T$ approaches $T_c$ from above, with most of the increase occurring between $2T_c$ and $T_c$. We will argue that such behavior is generic in a large class of confining theories with classical gravity duals. In a scan through various choices of $Z(\lambda)$, each of which reproduces the first two axial glueball mass ratios to within one sigma, we find that the increase can be as large as $60\%$. For our optimal choice of $Z(\lambda)$, which provides the best fit to the lattice results for the first two axial glueball mass ratios, the increase is only $0.01\%$.

To obtain $\gcs$, we compute the low-frequency limit of $G_R(\omega,\vec{k}=0)$ holographically. In section~\ref{CorrCS} we initiate the study of $G_R(\omega,\vec{k})$ at non-zero $\omega$ and $|\vec{k}|$, in the $T\geq T_c$ regime. We focus in particular on $\textrm{Im}\,G_R(\omega,\vec{k})$, which is proportional to the spectral function of $q(x^{\mu})$. After suitably subtracting the high-frequency asymptotics, by computing
the difference in the value of the correlator at two temperatures, our results suggest the presence of a reasonably long-lived excitation with energy on order of the lightest axial glueball mass at $T=0$. That is sufficiently light to prompt the speculation that perhaps such an excitation could dominate many CP-odd phenomena in the QGP created in heavy ion collisions.

In section~\ref{discussion} we summarize our results and discuss directions for future research.

\section{Improved Holographic QCD}
\label{review}

The holographic model that we consider as the dual to pure large-$N_c$ YM is (4+1)-dimensional Einstein-dilaton gravity with a well-chosen dilaton potential~\cite{Gursoy:2007cb,Gursoy:2007er,Gursoy:2008bu,Gursoy:2008za,Kiritsis:2009hu,Gursoy:2009jd,Gursoy:2009kk,Gursoy:2010fj}. In terms of the holographic 't Hooft coupling $\lambda \equiv e^{\Phi}$, the bulk action is
\be
S = M_p^3 N_c^2 \int d^5 x \sqrt{-g} \left[ R  -{4\over 3} \frac{(\partial \lambda)^2}{\lambda^2} +V(\lambda)\right]+S_{\textrm{bdry}},
\label{1}
\ee
where $M_p$ is the Planck Mass, related to the (4+1)-dimensional Newton's constant $G_5$ as $M_p^3={1/(16 \pi G_5 N_c^2)}$, $g$ and $R$ are the determinant and Ricci scalar of the bulk metric, $V(\lambda)$ is the dilaton potential, and $S_{\textrm{bdry}}$ represents all boundary terms, including the Gibbons-Hawking term as well as the counterterms needed for holographic renormalization\cite{Papadimitriou:2011qb}.

If $V(\lambda)=12/\ell^2$ with a constant length scale $\ell$, then the equations of motion arising from eq.~\eqref{1} admit a solution with constant $\lambda$ and an AdS metric with radius of curvature $\ell$,
\be
\label{adsmetric}
ds^2_{AdS} = \frac{\ell^2}{r^2} \left(dr^2 - dt^2 + d\vec{x}^2 \right), \qquad 0 < r < \infty.
\ee
Here $r$ is the holographic radial coordinate, dual to the field theory energy scale: the region near the AdS boundary at $r\to0$ is dual to the ultra-violet (UV) of the field theory, while the region near the Poincar\'{e} horizon at $r\to\infty$ is dual to the infra-red (IR). Such a solution describes a conformal field theory.

For non-trivial $V(\lambda)$, the equations of motion admit vacuum solutions in which $\lambda$ depends only on $r$ and the metric takes the form
\be
\label{eq:zerotsol}
ds^2= b_0(r)^2 (dr^2 -dt^2+ d\vec{x}^2), \qquad 0<r<\infty,
\ee
with warp factor $b_0(r)$. In IHQCD we demand that as $r\to0$ the metric approach that of AdS, $b_0(r) \to \ell/r$ (up to corrections logarithmic in $r$) and that $\lambda$ vanish logarithmically, $\lambda \to -1/\log r$, to mimic the running of the large-$N_c$ YM coupling.

Large-$N_c$ YM approaches a free theory in the UV, so we expect the holographic dual in the $r\to0$ region to be a string theory, not just a classical gravity theory like IHQCD. On the other hand, in large-$N_c$ YM, $\lambda_t$ diverges in the IR, so a classical gravity theory may be a reliable description in the $r\to\infty$ region. IHQCD is intended to be such a \textit{low-energy effective} description of large-$N_c$ YM, reliable in the $r\to\infty$ region. In practice, the role of the $r\to0$ region in IHQCD is simply to provide boundary conditions for the fields in the $r\to\infty$ region. We impose those boundary conditions at a cutoff, \textit{i.e.}~at some small but finite $r=\epsilon$. We then compute low-energy quantities that are insensitive to the cutoff, some of which we match to large-$N_c$ YM, while the rest are predictions of the model. A more detailed discussion of these issues appears for example in ref.~\cite{Kiritsis:2009hu}.

Using classical gravity in the $r\to0$ region has an important consequence, however: IHQCD will actually be dual to a theory that flows to a \textit{non-trivial} UV fixed point. Generically, the UV physics of large-$N_c$ YM and IHQCD will thus be different. For example, in IHQCD, $\eta/s=1/(4\pi)$~\cite{Kovtun:2004de}, which is much smaller than the high-$T$ perturbative result for $\eta/s$ in large-$N_c$ YM~\cite{Arnold:2000dr}. Nevertheless, in order to match IHQCD to known results for IR quantities in large-$N_c$ YM, we must match some quantities in the UV. For example, to reproduce lattice results for the free energy of large-$N_c$ YM for $T\gtrsim T_c$ with the correct normalization, we must demand that at high $T$ the free energy of IHQCD obey a Stefan-Boltzmann law. That requirement fixes the value of $\ell$ in the asymptotic AdS region in units of the Planck mass: $(M_p \ell)^{-3} = 45\pi^2$~\cite{Gursoy:2008bu,Gursoy:2008za}.

By matching to another UV quantity, the perturbative large-$N_c$ YM $\beta$-function, we can also constrain $V(\lambda)$. In the $r\to0$ region, where $\lambda$ is small, $V(\lambda)$ has a regular series expansion
\be
\label{vsmall}
V(\lambda)={12 \over \ell^2} \left(1+ v_0 \l +v_1 \l^2+ \mathcal{O}\left(\lambda^3\right)\right).
\ee
Committing to an identification of the field theory renormalization scale $E\equiv E_0 \, b_0(r)$, where $E_0$ can be fixed by matching to the lowest glueball mass or to the result for $T_c$ from lattice large-$N_c$ YM, we can fix the coefficients $v_0$ and $v_1$ in terms of the coefficients of the perturbative large-$N_c$ YM $\beta$-function~\cite{Gursoy:2007cb,Gursoy:2007er,Gursoy:2008za,Gursoy:2010fj,Papadimitriou:2011qb}:
\begin{subequations}
\be
\beta(\lambda_t) = -\beta_0 \lambda_t^2 -\beta_1 \lambda_t^3 + \cO(\lambda_t^4), \qquad \beta_0=\frac{22}{3(4\pi)^2}, \qquad \beta_1=\frac{51}{121}\beta_0^2,
\ee
\be
\label{eq:betamatch}
v_0 = \frac{8}{9}\beta_0, \qquad v_1={4\over 9}\beta_1 + {23\over 81}\beta_0^2.
\ee
\end{subequations}

In the vacuum solutions, generically $\lambda$ diverges as $r\to\infty$. The large-$\lambda$ expansion of $V(\lambda)$ must take the form $V(\lambda)\propto \lambda^{4\over 3}\sqrt{\log \lambda}$ in order for the glueball spectrum to be gapped and discrete with asymptotically linear trajectories~\cite{Gursoy:2007cb,Gursoy:2007er,Kiritsis:2009hu,Gursoy:2010fj}. With this asymptotic form for $V(\lambda)$, as $r\to\infty$ the warp factor and $\lambda(r)$ take the form
\be
\label{rlargemetric}
b_0(r)\propto e^{-\left(r/L\right)^2}, \qquad \lambda(r)\propto \frac{r}{L}~e^{\frac{3}{2}\left(r/L\right)^2}, \qquad (r\to\infty)
\ee
where $L$ is a length scale determined by the value of $\lambda$ at $r=\epsilon$. The form of $b_0(r)$ in eq.~\eqref{rlargemetric} is sufficient to guarantee that the dual field theory is confining~\cite{Gursoy:2007cb,Gursoy:2007er}. The metric actually has a mild singularity\footnote{On the other hand, in the string frame, where the metric scale factor is $\lambda^{2/3}(r)b_0(r)$, the curvature approaches zero as $r\to\infty$.} at $r= \infty$ that can be cloaked by a regular horizon and hence is a ``good'' singularity~\cite{Gubser:2000nd}. Moreover, the singularity is repulsive~\cite{Gursoy:2007er,Gursoy:2008za}, which guarantees that the low-energy spectrum and other observables are insensitive to the details of the resolution of the singularity.

\begin{figure}[h!]
 \begin{center}
\includegraphics[scale=1.5]{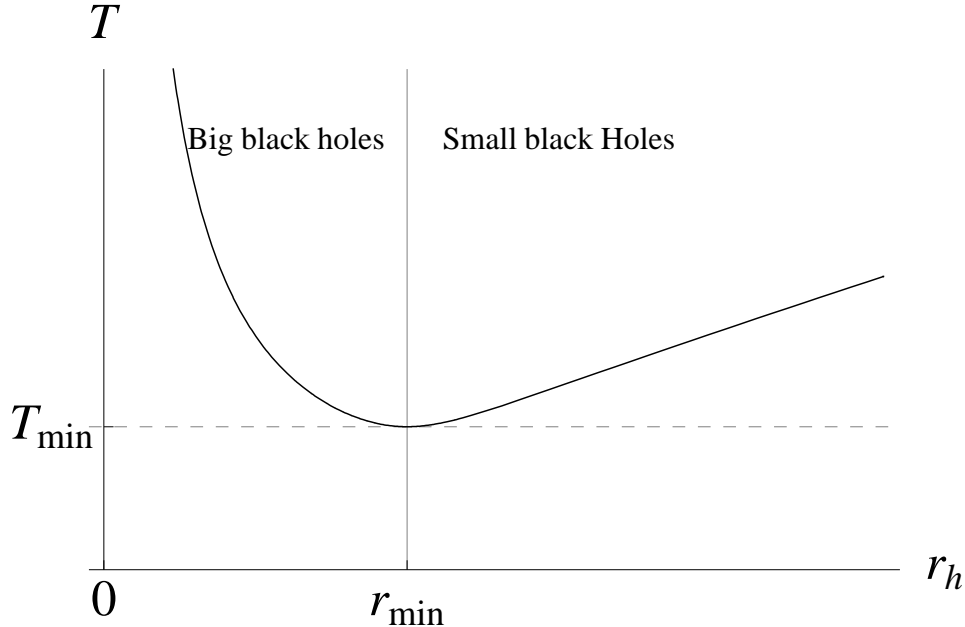}\\
 \end{center}
 \caption[]{Schematic plot for the typical form of the black hole Hawking temperature $T$ as a function of the horizon position $r_h$, for a generic choice of $V(\lambda)$ (with the correct small- and large-$\lambda$ asymptotics). The temperature exhibits a minimum, $T_{min}$, at $r_{\textrm{min}}$, which separates the large black hole ($r_h<r_{\textrm{min}}$) from the small black hole ($r_h>r_{\textrm{min}}$) branches.
}
\label{bhn}
\end{figure}

The black hole solutions of the model defined by the action in eq.~\eqref{1} have non-trivial $\lambda(r)$ and a metric of the form~\cite{Gursoy:2008bu,Gursoy:2008za}
\be\label{bh}
ds^2=b(r)^2\left({dr^2 \over f(r)}-f(r) dt^2 + d\vec{x}^2\right), \qquad 0<r < r_h.
\ee
The surface $r=r_h$ is the horizon, where $f(r_h)=0$, and the  corresponding Hawking temperature is $T=4\pi f'(r_h)$. Black hole solutions only exist for temperatures above a value $T_{\textrm{min}}$, and in fact two branches of solutions exist, the large and small black holes (comparing $r_h$ to $\ell$). Fig.~\ref{bhn} depicts the typical form of $T$ as a function of $r_h$, including the two branches of black hole solutions. For both large and small black holes, as $r\to r_h$, the warp factor $b(r)$ asymptotes to a constant whose value determines the entropy density, $s=b(r_h)^3/(4G_5)$, and as $r\to0$, $b(r)\to r/\ell$, up to ${\cal O}(r^4)$ (times logarithmic) corrections, indicating that in the field theory the thermal energy density and pressure are both of order $N_c^2$.

In large black hole solutions, $\lambda(r)$ decreases monotonically as $T$ increases, so that $\lambda \to 0$ as $T\to\infty$. In small black hole solutions, $\lambda(r)$ increases as $T$ increases. In particular, as discussed in refs.~\cite{Gursoy:2008za,Gursoy:2009jd}, the value of $\lambda(r)$ at the horizon, $\lambda_h\equiv \lambda(r_h)$, is a monotonically increasing function of $r_h$, so a plot of $T$ versus $\lambda_h$ is qualitatively similar to fig.~\ref{bhn}: in the $T\to\infty$ limit, $\lambda_h \to 0$ on the large black hole branch and $\lambda_h \to \infty$ on the small black hole branch (see for example fig. 2 (a) of ref.~\cite{Gursoy:2008za}).

If we Wick-rotate to a compact Euclidean time direction of length $1/T$, then for $T\geq T_{min}$ three bulk solutions exist: the Wick-rotated version of eq.~\eqref{eq:zerotsol}, which describes a thermal gas of gravitons and is dual to a confined state, and the Wick-rotated large and small black holes, which are dual to deconfined states. To determine which solution is thermodynamically preferred at any given $T$, we must determine which has the smallest on-shell Euclidean action, dual to the field theory's free energy (times $1/T$). As shown in refs.~\cite{Gursoy:2008bu,Gursoy:2008za,Gursoy:2009jd}, the small black hole solutions are never thermodynamically preferred, but at some $T_c>T_{min}$ the large black hole solutions become thermodynamically preferred. Indeed, the system exhibits a first-order Hawking-Page type transition at $T_c$, dual to a confinement-deconfinement transition.

In general, for a given potential $V(\lambda)$ we cannot solve the equations of motion arising from eq.~\eqref{1} exactly, so we resort to numerics. Here we will only sketch our numerical procedure, which is described in detail for example in ref.~\cite{Gursoy:2009jd}. At the cutoff $r=\epsilon$ we impose a Dirichlet condition on each field, and in particular we demand that the metric take the AdS form. We then fix the remaining integration constants, including $\lambda_h$, by a shooting algorithm. Given a choice of $V(\lambda)$ and the Dirichlet conditions at $r=\epsilon$, we obtain a one-parameter family of solutions labeled by $T$, or equivalently by $\lambda_h$. Following refs.~\cite{Gursoy:2009jd,Gursoy:2009kk}, in our numerics we use a simple form for $V(\lambda)$ with the correct small- and large-$\lambda$ asymptotics,
\be
\label{vform}
V(\lambda) = \frac{12}{\ell^2} \left[ 1+V_0\lambda + V_1 \lambda^{4/3} \sqrt{\log\left(1+ V_2 \lambda^{4/3} + V_3 \lambda^{2}\right)}\right].
\ee
Expanding eq.~\eqref{vform} about $\lambda=0$ and matching to eq.~\eqref{vsmall}, we find $v_0=V_0$ and $v_1=V_1\sqrt{V_2}$. The coefficients $V_0$ and $V_2$ can be determined in terms  of $V_1$ by imposing  the conditions in eq.~\eqref{eq:betamatch}. The potential thus has two free parameters, $V_1$ and $V_3$. We fit these two parameters by matching to lattice results for two thermodynamic quantities in large-$N_c$ YM: the latent heat of the deconfinement transition, which is proportional to the entropy density at the transition, $s(T_c)/(N_c^2T_c^3) \simeq 0.31$ \cite{Lucini:2005vg}, and the pressure at $T=2T_c$~\cite{Boyd:1996bx,Lucini:2005vg,Bringoltz:2005rr}. Upon fixing  $V_1$ and $V_3$ in this fashion, IHQCD describes very well both the $T=0$ glueball spectra ($0^{++}$ and $2^{++}$) as well as the finite $T$ thermodynamics of large-$N_c$ YM~\cite{Gursoy:2009jd,Panero:2009tv}.

The instanton number density operator, $q(x^{\mu})$ in eq.~\eqref{q}, is dual to a bulk pseudoscalar, the axion $\alpha$ (as in many top-down models). In the field theory, the source for $q(x^{\mu})$ is an angular variable, the $\theta$-angle. As a result, the action of the bulk axion must be invariant under shifts of $\alpha$, and hence must depend only on derivatives\footnote{Instanton effects may produce a non-trivial axion potential, such as a term $\cos \alpha$. These instanton effects are exponentially suppressed in the large-$N_c$ limit, however.} $\partial \alpha$. General arguments in string theory and in YM theory, including the argument that $\theta$ dependence should appear in the YM vacuum energy only at order one in the large-$N_c$ limit rather than at order $N_c^2$~\cite{Witten:1998uka}, imply that the axion action is suppressed by ${\mathcal O}({1/N_c^2})$ compared to the action $S$ in eq.~\eqref{1} \cite{Gursoy:2007cb,Gursoy:2007er,Kiritsis:2009hu,Gursoy:2010fj}. We thus add to the model an axion with an action $S_{\alpha}$ of the form~\cite{Gursoy:2007cb,Gursoy:2007er,Kiritsis:2009hu,Gursoy:2010fj}
\be\label{axionaction}
S_{\alpha} = -\frac{1}{2}M_p^3 \int d^5x \sqrt{-g} ~Z(\lambda) (\partial \a)^2,
\ee
where, following the rules of effective field theory, we have included a dimensionless, $\lambda$-dependent normalization function, $Z(\lambda)$, consistent with the symmetries.

Being a massless pseudo-scalar, in an expansion of $\alpha(r)$ about $r=0$, the leading, non-normalizable term is a constant, which is proportional to the YM $\theta$-angle defined in the UV,
\be
\label{k}
\alpha(r=0)=\kappa\,\theta,
\ee
where in top-down models the proportionality constant $\kappa$ will be fixed, but not in bottom-up models. In other words, in our model  the normalization of the operator dual to $\alpha$ is ambiguous: $\alpha$ is dual to $q(x^{\mu})/\kappa$. Nevertheless, by fixing the normalization of the topological susceptibility we will be able to compute two-point functions of $q(x^{\mu})$ unambiguously, as we explain below.

To specify $S_{\alpha}$ completely we must specify $Z(\lambda)$. In principle, $Z(\lambda)$ can be fixed as follows. First, perform a lattice calculation of the Euclidean two-point function of $q(x^{\mu})$ with non-zero $T$ for some set of frequencies. Second, compute the same Euclidean two-point function holographically for all frequencies for some choice of $Z(\lambda)$. A least squares fit of the holographic results to the lattice results should then determine $Z(\lambda)$. To study generic behavior of holographic models, we will instead proceed by using simple forms for $Z(\lambda)$ that we constrain by matching to lattice results for the topological susceptibility and axial glueball mass spectrum. Notice that matching to any lattice data will always have room for improvement: lattice definitions of $q(x^{\mu})$ generically suffer from power-law divergences that dominate in the continuum limit, making lattice calculations of correlators of $q(x^{\mu})$ noisy~\cite{Vicari:2008jw}. Accurate calculations may be possible in the near future\footnote{We thank F.~Bruckmann, H.~Panagopoulos, and A.~Sch\"{a}fer for discussions on this issue.}.

We can constrain $Z(\lambda)$ as follows. Since $Z(\lambda)$ is the coefficient of a kinetic term, we demand that $Z(\lambda)\geq0$. We can also constrain $Z(\lambda)$'s small- and large-$\lambda$ asymptotics~\cite{Gursoy:2007cb,Gursoy:2007er,Kiritsis:2009hu,Gursoy:2010fj}:
\be
Z(\lambda) \propto \begin{cases} Z_0 + {\cal O}(\lambda), & \lambda \to 0, \\ \lambda^4 + {\cal O}(1/\lambda), & \lambda \to \infty, \end{cases}
\ee
where $Z_0$ is a dimensionless constant. The small-$\lambda$ form follows from the rules of effective field theory: a constant is the most general allowed term. The large $\lambda$ behavior is fixed by glueball universality~\cite{Gursoy:2007er}. Various towers of glueballs have linear asymptotic trajectories: for large excitation number $n$, their squared masses go as $(m^i_{n})^2 = c^i n+\cdots$, with constants $c^i$, where the integer $i$ labels different towers. Glueball universality is the statement that all the slopes $c^i$ are similar, \textit{i.e.}\ do not depend on $i$. That is automatic for the $0^{++}$ and $2^{++}$ glueballs. Requiring the same for the $0^{-+}$ glueballs forces $Z(\lambda)$ to go as $\lambda^4$ at large $\lambda$~\cite{Gursoy:2007er}.

We will use the simplest form of $Z(\lambda)$, also used for example in ref.~\cite{Gursoy:2009jd},
\be
\lab{Zspec}
Z(\lambda)=Z_0 (1+c_4 \lambda^4),
\ee
where $c_4$ is a dimensionless constant. To fix $Z_0$ we match to the large-$N_c$ YM lattice result for the Euclidean topological susceptibility, $\chi$, defined in terms of the $T=0$ vacuum energy density ${\mathcal{E}}(\theta)$ as
\be
\chi \equiv  {d^2\E(\theta)\over d\theta^2}=\int d^4x \left\langle q(x^{\mu}) q(0)\right\rangle_{\textrm{E}},
\ee
where the subscript E denotes the Euclidean correlator. The holographic result for $\chi$ is\footnote{The holographic calculation of $\chi$ in ref.~\cite{Gursoy:2007er} assumed $\kappa=1$. Here we allow for arbitrary $\kappa$.}~\cite{Gursoy:2007er},
\be
\chi=\frac{\kappa^2 M_p^3}{\int_0^{\infty}\frac{dr}{b_0^{3}(r) Z(\lambda(r))}}.
\label{magsus}
\ee
Clearly $\chi$ will be proportional to $\kappa^2 Z_0$. Thus, for any given value of the parameter $c_4$,  matching the holographic result for $\chi$ to the lattice result, $\chi \approx (191\,\textrm{MeV})^4$~\cite{DelDebbio:2004ns,Vicari:2008jw}, fixes the product $\kappa^2 Z_0$. On the other hand, since the locations of poles in the two-point function of $q(x^{\mu})$ are independent of the overall normalization $\kappa^2Z_0$, we can fix $c_4$ independently  by matching the mass of the lowest $0^{-+}$ glueball to the lattice result of ref.~\cite{Morningstar:1999rf}\footnote{For a recent lattice study of the glueball spectrum at large $N$, see ref.~\cite{Lucini:2010nv}. We prefer to use the older results of ref.~\cite{Morningstar:1999rf} because in the latter work an excited state of the $0^{-+}$ tower is given.},
\be
\label{lowestaxial}
m_{0^{-+}}/m_{0^{++}} = 1.50(4).
\ee
The resulting values are\footnote{The result for $\kappa^2 Z_0$ in ref.~\cite{Gursoy:2009jd} was too large by a factor of four, producing an erroneous result, $\kappa^2 Z_0=133$. In eq.~\eqref{eq:Zparams} we present the correct value, $\kappa^2 Z_0 = 133/4 = 33.25$.} \cite{Gursoy:2009jd}
\be
\label{eq:Zparams}
\kappa^2Z_0=33.25, \qquad c_4=0.26.
\ee
These values can then be used to predict the masses in the full tower of $0^{-+}$ glueballs. As shown in refs.~\cite{Gursoy:2007cb,Gursoy:2007er,Kiritsis:2009hu,Gursoy:2010fj}, the holographic result for the first excited $0^{-+}$ glueball mass, $m_{0^{*-+}}$, agrees very well with the lattice result~\cite{Morningstar:1999rf},
\be
\label{firstexcitedaxial}
m_{0^{*-+}}/m_{0^{++}} = 2.11(6).
\ee

Crucially, notice that by fixing the normalization of $\chi$ we have fixed the normalization of {\it any} two-point function of $q(x^{\mu})$, and thus have eliminated the normalization ambiguity mentioned below eq.~\eqref{k}. In other words, the holographic calculation of the two-point functions of $q(x^{\mu})$ will only depend on the combination $\kappa^2Z_0$ (as we will see explcitly in section~\ref{GammaCS}), which we have fixed to the value in eq. (\ref{eq:Zparams}).

\begin{figure}[h]
 \begin{center}
\includegraphics[scale=1]{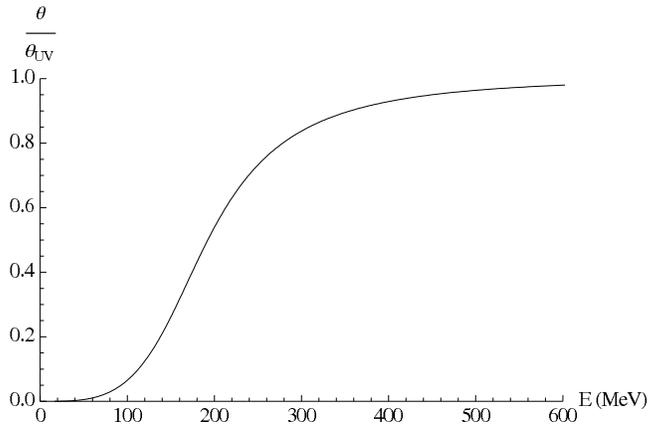}
\end{center}
 \caption[]{The normalizable solution $\alpha(r)$ for the axion at $T=0$, expressed as a running $\theta$-angle normalized to the UV value, as a function of the energy scale, $E(r)=E_0 b(r)$. We fix $E_0$ by matching our holographic result for $T_c$ to the large-$N_c$ YM lattice result.}
\label{fig1}
\end{figure}

Solutions for $\alpha$ as a function of $T$ were studied in refs.~\cite{Gursoy:2008za,Gursoy:2009jd}. When $T=0$, a non-trivial UV $\theta$-angle forces $\alpha(r)$ to be non-trivial. The resulting normalizable solution then indicates that the non-zero UV $\theta$-angle flows to zero in the IR, as shown in fig.~\ref{fig1}, and additionally triggers a non-zero $\langle q(x)\rangle/\kappa$. The $T=0$ solution for $\alpha(r)$ is unchanged when $T<T_c$: Wick-rotating the metric in eq.~\eqref{eq:zerotsol} to a compact Euclidean time does not affect the static solution $\alpha(r)$. Such behavior is expected in a confined phase at leading order in $N_c$, due to large-$N_c$ volume independence. When $T>T_c$, however, the only non-singular solution for the axion is a constant, $\alpha(r) = \kappa \,\theta$, indicating that $\langle q(x)\rangle=0$, in agreement with evidence from lattice data for large-$N_c$ YM~\cite{Vicari:2008jw}.

What is the behavior of the topological susceptibility as a function of temperature, $\chi(T)$? When $T<T_c$, $\chi(T)$ is independent of $T$, \textit{i.e.}\ takes the same value as at $T=0$, eq.~\eqref{magsus}, again due to large-$N_c$ volume independence. When $T>T_c$, the holographic result for the topological susceptibility is
\be
\chi(T)={\kappa^2 M_p^3 \over \int_0^{r_h} {dr \over b^{3}(r) f(r) Z(\l(r))}}. \qquad (T>T_c)
\label{magsus1}
\ee
The denominator on the right-hand-side of eq.~\eqref{magsus1} diverges at the black hole horizon, so in fact $\chi(T)=0$ when $T>T_c$, up to ${\cal{O}}(e^{-N_c})$ corrections~\cite{Gursoy:2008za,Gursoy:2009jd}.

\begin{figure}[t]
 \begin{center}
\includegraphics[scale=1.5]{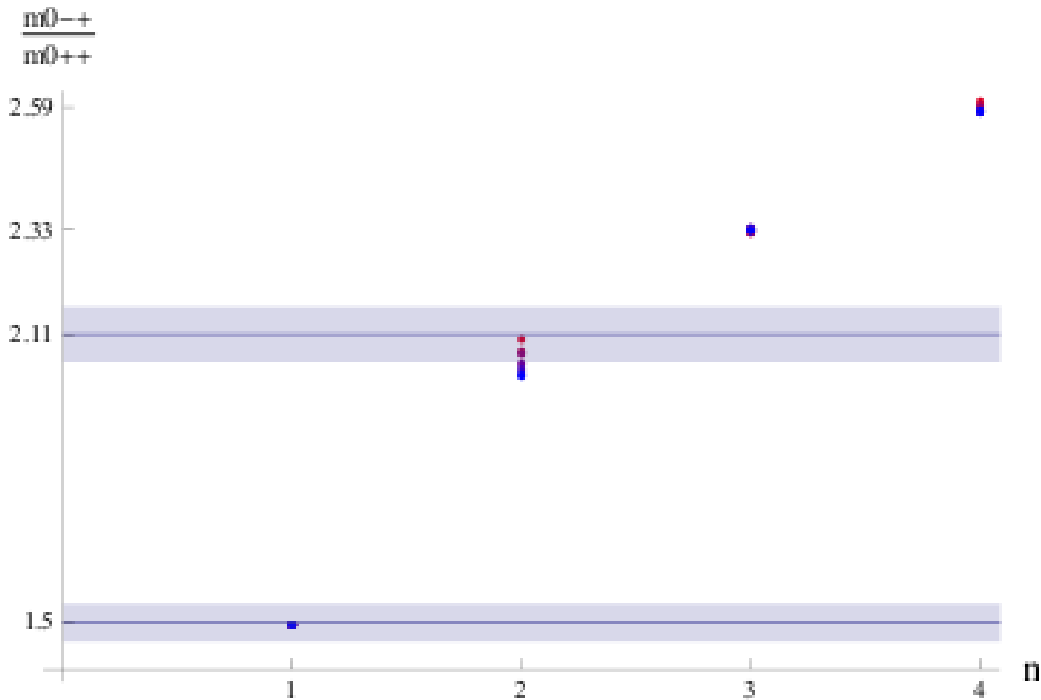}
\end{center}
 \caption[]{Our holographic results for the masses of the $0^{-+}$ glueball states with excitation number $n$, normalized to the lowest $0^{++}$ glueball mass, obtained by varying the coefficients $c_1$ and $c_4$ in the $Z(\lambda)$ in eq.~\eqref{c1c4}. From the top (red) dots (visible only for $n=2$ and $n=4$) to the bottom (blue) dots, $(c_1,c_4) = (0,0.26), (0.5, 0.87), (1, 2.2), (5, 24), (10, 75), (20, 230), (40, 600)$. The lowest axial glueball mass, $n=1$, is always fixed to be the value in eq.~\eqref{lowestaxial}. The two horizontal blue lines with surrounding blue bands indicate the results and errors, respectively, of the large-$N_c$ YM lattice calculations for the masses of the lowest and first excited states, $n=1$ and $n=2$ (see eqs.~\eqref{lowestaxial} and~\eqref{firstexcitedaxial})~\cite{Morningstar:1999rf}. Only the mass of the $n=2$ state is appreciably sensitive to changes of $c_1$ and $c_4$, differing from the lattice result by $3\%$ at most. }
\label{figaxialmass}
\end{figure}

We will also consider a form for $Z(\lambda)$ more general than that of eq.~\eqref{Zspec}. On the large black hole branch, if $T$ is large then $\lambda$ is small, in which case we expect the largest polynomial correction to the $Z(\lambda)$ in eq.~\eqref{Zspec} to be a term linear in $\lambda$, hence we consider
\be\label{c1c4}
Z(\lambda) = Z_0\left(1 + c_1 \lambda + c_4 \lambda^4\right),
\ee
where $c_1$ is a dimensionless constant, which we choose to be positive. If we continue to fit only to the lattice result for the lowest $0^{-+}$ glueball mass, we find a substantial degeneracy (which is not surprising, given that we have introduced an additional parameter, $c_1$). Specifically, for any positive value of $c_1$, a value of $c_4$ exists such that, upon matching to the lowest axial glueball mass in eq.~\eqref{lowestaxial}, the value of the first excited axial glueball mass is in rough agreement with the value in eq.~\eqref{firstexcitedaxial}, exhibiting at most a $3\%$ discrepancy, as shown in fig.~\ref{figaxialmass}. To constrain $c_1$ we will demand that our holographic results for the axial glueball masses fall within one sigma of the lattice values in eqs.~\eqref{lowestaxial} and~\eqref{firstexcitedaxial}. That results in the constraints
\be
\label{onesigmalimits}
0\lesssim c_1 \lesssim 5, \qquad 0.06\lesssim c_4 \lesssim 50.
\ee
In fact, the optimal values, which provide the best fit, are the ones in eqs.~\eqref{Zspec} and~\eqref{eq:Zparams}: $(c_1,c_4)=(0,0.26)$.

As we have seen, the function $Z(\lambda)$ must be non-negative and is constrained in the $\lambda\to0$ and $\lambda\to \infty$ limits. For intermediate values of $\lambda$, the most natural assumption is that $Z(\lambda)$ is monotonic. At least, we are not aware of any compelling evidence for the existence of maxima or minima in $Z(\lambda)$. Our choices for $Z(\lambda)$ were thus monotonic functions of $\lambda$, namely polynomials in $\lambda$ with strictly positive coefficients. To test the effect of maxima and minima in $Z(\lambda)$, we considered two changes to the $Z(\lambda)$ in eq.~\eqref{c1c4}. First, we allowed slightly negative $c_1$, while maintaining $Z(\lambda)\geq0$. Second, we introduced a maximum by a adding a Gaussian peak to $Z(\lambda)$. In each case we computed the axial glueball mass spectrum. After matching to the lattice result for the lowest axial glueball mass, we found that the fit to the first excited axial glueball mass was worse, deviating from the lattice result by about $10\%$. We consider that a preliminary indication that monotonic $Z(\lambda)$ may indeed be the best choice. We leave more thorough tests for future research.

Our assumption that $Z(\lambda)$ is monotonic in $\lambda$ determines the qualitative behavior of $Z(\lambda)$ as a function of $T$. On the large black hole branch, as $T\to\infty$, $\lambda\to0$, and as $T$ decreases towards $T_c$, $\lambda$ increases monotonically. As a result, for our choices of $Z(\lambda)$---simple polynomials in $\lambda$ with positive coefficients---when $T\to\infty$, $Z(\lambda)\to Z_0$, and when $T \to T_c$, $Z(\lambda)$ will increase monotonically. As functions of $T$, our $Z(\lambda)$ are thus bounded from below by their value in the $T\to\infty$ limit: $Z(\lambda) \geq Z_0$. The behavior of $Z(\lambda)$ as a function of $T$ will translate directly into the behavior of $\gcs$ as a function of $T$, as we will show in the next section. In particular, the dimensionless combination $\gcs/(sT)$ will be bounded from below by its value in the $T\to\infty$ limit, and will increase as $T\to T_c$. In the next section we will also present a more general argument that $\gcs/(sT)$ must increase as $T$ approaches $T_c$ from above.

\section{The Chern-Simons Diffusion Rate}
\lab{GammaCS}

We will compute $\gcs$ using eq.~\eqref{gcs0}, rewritten as
\be
\label{gcs}
\Gamma_{CS}=-\k^2~\lim_{\omega \to 0}\,\frac{2T}{\omega}\,\textrm{Im}\, \hat{G}_R (\omega,\vec{k}=0),
\ee
where $\hat{G}_R(\omega,\vec{k})$ is the retarded two-point function of $q(x^{\mu})/\kappa$, the operator dual to our axion $\alpha$.

In holography, the on-shell bulk action is the generating functional for field theory correlation functions~\cite{Gubser:1998bc,Witten:1998qj}. To compute the two-point function $\hat{G}_R(\omega,\vec{k})$ in the high-temperature, deconfined phase of IHQCD, we must solve the linearized equation of motion of the axion in the black hole spacetime with metric in eq.~\eqref{bh}, with $T\geq T_c$. We thus introduce a fluctuation of the axion, $\delta \alpha(r,x^{\mu})$, where $x^{\mu} = (t,\vec{x})$. When $T\geq T_c$, the background solution for the axion is trivial, hence the linearized equation of motion for $\delta\alpha(r,x^{\mu})$ is simply
\be
\label{fluceq1}
\frac{1}{Z(\lambda(r)) \sqrt{-g}} \,\partial_r \left[Z(\lambda(r)) \sqrt{-g} \, g^{r r}\partial_r \delta \alpha(r,x^{\mu})\right]+g^{\mu \nu} \partial_{\mu} \partial_{\nu} \,\delta \alpha(r,x^{\mu}) = 0,
\ee
where the metric is that of eq.~\eqref{bh}. Notice in particular that $\delta \alpha$ will not couple to the fluctuations of any other fields because the background solution preserves CP and the axion is the only CP-odd field in the bulk. We must solve eq.~\eqref{fluceq1} with Dirichlet boundary condition at the asymptotically AdS boundary and with in-going wave boundary condition at the horizon~\cite{Son:2002sd}. The solution takes the form
\be
\delta \a (r,x^{\mu}) = \int {d^4 k \over (2 \pi)^4} \, e^{ik x} \, \delta\alpha(r,k^{\mu})\,a(k^{\mu}),
\ee
where $k^{\mu}=(\omega,\vec{k})$ and where $a(k^{\mu})$ is fixed by the Dirichlet boundary condition,
\be
\lim_{r \to 0} \delta \alpha(r,x^{\mu}) = \int {d^4 k \over (2 \pi)^4} \, e^{ik x} \, a(k^{\mu}),
\ee
while $\delta \alpha(r,k^{\mu})$ obeys the equation
\be
\label{fluceq}
\frac{1}{Z(\lambda(r)) \sqrt{-g}} \,\partial_r \left[Z(\lambda(r)) \sqrt{-g} \, g^{r r}\partial_r \delta \alpha(r,k^{\mu})\right]-g^{\mu \nu} k_{\mu} k_{\nu} \,\delta \alpha(r,k^{\mu}) = 0,
\ee
with unit normalization at the asymptotically AdS boundary, $\lim_{r\to 0} \delta \alpha(r,k^{\mu}) = 1$, and in-going wave boundary condition at the horizon. The on-shell axion action is then
\be
\label{onshact}
S^{\textrm{on-shell}}_{\alpha}=\int \left.  {d^4 k \over (2 \pi)^4} \,a(-k^{\mu})\,{\mathcal{F}}(r,k^{\mu}) \, a(k^{\mu})\right| _0^{r_h},
\ee
where
\be
\label{Fdef}
{\mathcal{F}}(r,k^{\mu})\equiv -{M_p^3 \over 2} \, \delta\alpha(r,-k^{\mu}) \, Z(\lambda(r))\,\sqrt{-g} \, g^{rr} \,\partial_r  \delta\alpha(r,k^{\mu}).
\ee
The retarded Green's function is then~\cite{Son:2002sd}
\be
\label{grf}
\hat{G}_{R}(\omega,\vec{k})= -2 \, \lim_{r\to 0} {\mathcal{F}}(r,k^{\mu}).
\ee

To compute $\gcs$, we need to solve eq.~\eqref{fluceq} with $\vec{k}=0$ and with small $\omega$. We will do so in two ways, first using near-horizon matching and second using the membrane paradigm, following ref.~\cite{Iqbal:2008by}. In each case we can determine $\gcs$ analytically, essentially because $\delta \alpha$ is a massless fluctuation.

In the near-horizon matching technique, we first solve eq.~\eqref{fluceq} with $\omega=0$ and then expand the solution near the horizon. We then reverse the order of operations, solving the equation in the near-horizon region and then expanding the solution in $\omega$. Finally, we match the two solutions to obtain ${\mathcal{F}}(r,k^{\mu})$.

When $\vec{k}=0$ and $\omega=0$ the solution of eq.~\eqref{fluceq} is
\be
\label{fluczerofreq}
\delta \alpha=C_1+ C_2 \int_0^r \frac{dr'}{Z(\lambda(r'))b(r')^3f(r')},
\ee
with constant coefficients $C_1$ and $C_2$. The second term on the right-hand side of eq.~\eqref{fluczerofreq} diverges as $r \rightarrow r_h$. As a result, when $\omega=0$ a normalizable solution must have $C_2=0$. When $\omega$ is small but non-zero, a normalizable solution may have $C_2\propto \omega$. Plugging eq.~\eqref{fluczerofreq} into eq.~\eqref{Fdef}, we find
\be
\label{calFeq}
\lim_{r\to0}{\mathcal{F}}(r,k^{\mu}) = -\frac{M_p^3}{2} \, C_1 \, C_2. \qquad (\omega \ll T, \vec{k}=0)
\ee
We will choose $C_1=1$ so that our $\delta \alpha$ has unit normalization at the asymptotically AdS boundary. Our task is thus to determine $C_2$. Expanding the solution in eq.~\eqref{fluczerofreq} around the horizon, we find
\be
\label{solhor1}
\delta \alpha=C_1 + \frac{C_2}{Z (\lambda_h)\,b(r_h)^3\,f'(r_h)} \log(r_h-r) + \cO(r_h-r),
\ee
where $f'(r_h)=4\pi T$. Now we reverse the order of operations. Expanding eq.~\eqref{fluceq} in $(r_h-r)$, we find the solution in the near-horizon region,
\be
\label{flucnh}
\delta \alpha=C_+ (r_h -r)^{{i \omega\over 4 \pi T}} +C_- (r_h -r)^{-{i \omega\over 4 \pi T}},
\ee
with coefficients $C_{\pm}$ that depend on $\omega$ but not on $r$. We set $C_+=0$ so that the near-horizon solution is an in-going wave~\cite{Son:2002sd}. Now we expand the solution in eq.~\eqref{flucnh} for small $\omega$:
\be
\label{solhor2}
\delta \alpha=C_- - i \frac{\omega}{4 \pi T}\, C_-\, \log(r_h-r) + \cO(\omega^2/T^2).
\ee
By matching the constant and logarithmic terms in eqs.~\eqref{solhor1} and~\eqref{solhor2}, we find
\be
\label{match}
C_1=C_-, \qquad C_2 =  - i \omega \, Z(\lambda_h) \, b(r_h)^3 \, C_-.
\ee
Setting $C_1=1$, we obtain $\lim_{r\to0}{\mathcal{F}}(r,k^{\mu})$ via eq.~\eqref{calFeq} and then $\hat{G}_R(\omega,\vec{k})$ via eq.~\eqref{grf},
\be
\label{greensresult}
\hat{G}_R(\omega,\vec{k}=0) = -i \,\omega \, M_p^3 \, Z(\lambda_h) \, b(r_h)^3. \qquad (\omega \ll T)
\ee
We thus obtain our main result for $\gcs$,
\be
\lab{gcs1}
\Gamma_{CS}=-\kappa^2\lim_{\omega \to 0}{2T \over \omega} \,\textrm{Im} \, \hat{G}_R(\omega,\vec{k}=0)= \frac{1}{N_c^2} \, {sT  \over 2 \pi}\,\kappa^2 Z(\lambda_h),
\ee
where we have used $M_p^3 = 1/(16\pi G_5 N_c^2)$ and where $s={b^{3}(r_h) \over 4 G_5}$ is the entropy density. Notice that the normalization of this result is fixed by the product $\kappa^2 Z_0$, which we fixed in section~\ref{review} by matching to the topological susceptibility at $T=0$.

The second equivalent, but more efficient, method that we will use to obtain $\hat{G}_R(\omega,\vec{k})$ is the membrane paradigm~\cite{Iqbal:2008by}. Kubo's formula for the retarded Green's function is
\be
\label{Kubo}
 \Pi(\omega,\vec{k}) =  \hat{G}_R(\omega,\vec{k}) \delta \alpha(\omega,\vec{k}),
\ee
where $\Pi(\omega,\vec{k})$ is the one-point function of $q(x^{\mu})/\kappa$ in Fourier space. Following ref.~\cite{Iqbal:2008by}, we extend eq.~\eqref{Kubo} into the bulk by defining an $r$-dependent response function,
\be
\zeta(r,\omega,\vec{k})\equiv{\Pi (r,\omega,\vec{k}) \over \omega \,M_p^3 \,\delta \alpha(r,\omega,\vec{k})},
\ee
where $\Pi(r,\omega,\vec{k})$ is the canonical momentum of $\delta\alpha(r,\omega,\vec{k})$ with respect to the $r$-foliation of the bulk space-time,
\be\lab{canmom}
\Pi(r,\omega,\vec{k})\equiv \frac{\delta S_{\alpha}}{\delta \partial_r \delta \alpha}= -M_p^3 \, Z(\lambda(r)) \sqrt{-g} \, g^{rr} \, \partial_r \delta \alpha (r,\omega,\vec{k}).
\ee
The retarded Green's function is then proportional to the boundary value of $\zeta$:
\be
\lab{Kubohol}
\hat{G}_R(\o,\vec{k}) = -M_p^3 \,\omega \lim_{r\to0} \zeta(r,\omega,\vec{k}).
\ee
An equation of motion for $\zeta$ is straightforward to derive using eq.~\eqref{canmom} and $\delta \alpha$'s equation of motion, eq.~\eqref{fluceq},
\bea
\label{zetaeq}
\partial_r \zeta & = & \frac{\omega}{Z(\lambda(r)) \sqrt{-g} \,g^{rr}} \left[\zeta^2+ Z(\lambda(r))^2 g \, g^{rr} g^{tt} \left(1+\frac{g^{xx}}{g^{tt}} \frac{\vec{k}^2}{\omega^2}\right) \right] \nonumber \\ & = & \frac{\omega}{Z(\lambda(r)) b(r)^3 f(r)} \left[\zeta^2+ Z(\lambda(r))^2 b(r)^6 \left(1-f(r) \frac{\vec{k}^2}{\omega^2}\right) \right].
\eea
To obtain the retarded Green's function $\hat{G}_R(\omega,\vec{k})$, we must impose regularity at the horizon, meaning $\partial_r \zeta$ is finite there~\cite{Iqbal:2008by}, hence the term in brackets in eq.~\eqref{zetaeq} must vanish\footnote{When $\vec{k}\neq 0$ but $\omega=0$, the boundary condition is modified from that in eq.~\eqref{bc}, as discussed in ref.~\cite{Iqbal:2008by}. In what follows, whenever we consider $\vec{k}\neq 0$ we will work with $\omega\neq 0$, hence we will use the boundary condition in eq.~\eqref{bc}.}
 at $r=r_h$:
\be
\label{bc}
\zeta(r_h)= +i Z(\lambda_h) b(r_h)^3.
\ee
We can now easily derive $\gcs$. In eq.~\eqref{zetaeq} we take $\vec{k}=0$ and observe that if $\omega \to 0$ then $\zeta$ becomes independent of $r$. The value of $\zeta$ for all $r$ is then the same as the value at the horizon, eq.~\eqref{bc}, and via eq.~\eqref{Kubohol} we trivially obtain $\hat{G}_R(\omega,\vec{k}=0)$, which is identical to eq.~\eqref{greensresult}. We thus find again
\be
\lab{gcs1-app}
\Gamma_{CS}=-\kappa^2\lim_{\omega \to 0}{2T \over \omega} \,\textrm{Im} \, \hat{G}_R(\omega,\vec{k}=0)= \frac{1}{N_c^2} \, {sT  \over 2 \pi}\,\kappa^2 Z(\lambda_h).
\ee

Our result suggests a natural dimensionless quantity to study,
\be
\frac{\Gamma_{CS}}{sT/N_c^2}=\frac{\kappa^2 Z(\lambda_h)}{2\pi},
\ee
which has implicit dependence on $T$ through $Z(\lambda_h)$, and is constant in $T$ if and only if $Z(\lambda)$ is a constant in $\lambda$, that is, if the axion does not couple to the dilaton. Indeed, as we mentioned at the end of section~\ref{review}, the behavior of $Z(\lambda)$ as a function of $T$ determines the behavior of $\Gamma_{CS}/(sT/N_c^2)$ as a function of $T$. In particular, on the large black hole branch, $\Gamma_{CS}/(sT/N_c^2)$ is bounded from below by its value in the $T\to\infty$ limit,
\be
\label{gcshighT}
\lim_{T\to\infty}\,\frac{\Gamma_{CS}}{sT/N_c^2} = \frac{\kappa^2 Z_0}{2\pi}.
\ee
If we use the preferred value $\kappa^2Z_0 = 33.25$~\cite{Gursoy:2009jd} then $\kappa^2Z_0/(2\pi) \simeq 5.29$. Moreover, $\Gamma_{CS}/(sT/N_c^2)$ will increase monotonically as $T$ approahces $T_c$ from above.

\begin{figure}[h]
 \begin{center}
\includegraphics[scale=1]{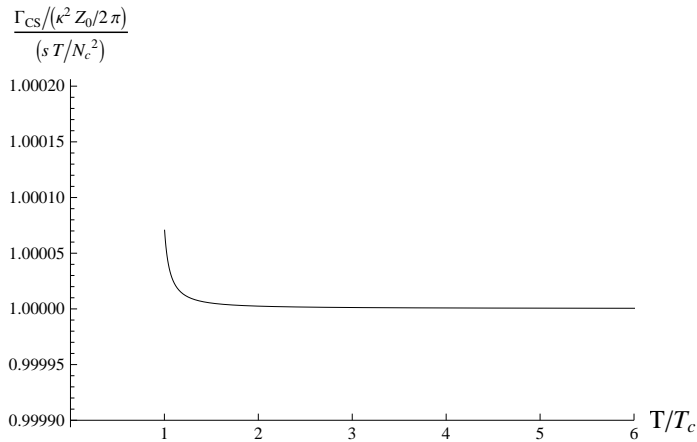}
 \end{center}
 \caption[]{Our numerical result for $\Gamma_{CS}/(sT/N_c^2)$, normalized to the $T\to\infty$ value $\kappa^2Z_0/(2\pi)$, as a function of $T/T_c$ for the $Z(\lambda)$ given in eq.~\eqref{Zspec}, with $c_4=0.26$~\cite{Gursoy:2009jd}. As $T$ decreases, $\Gamma_{CS}/(sT/N_c^2)$ remains nearly constant, experiencing only an approximately $0.01\%$ increase, mostly between $2T_c$ and $T_c$.}
\label{fig3}
\end{figure}

For the simplest choice of $Z(\lambda)$, given in eq.~\eqref{Zspec}, $\Gamma_{CS}/(sT/N_c^2)$ has extremely mild dependence on $T$: as $T$ approaches $T_c$ from above, $\Gamma_{CS}/(sT/N_c^2)$ is nearly constant, experiencing an increase of only about $0.01\%$, mostly between $2 T_c$ and $T_c$, as shown in fig.~\ref{fig3}. In bulk terms, the reason for this mild $T$ dependence is that between $T\to\infty$ and $T=T_c$, $\lambda_h$ increases from zero up to only $\lambda_h \approx 0.14$, which for the $Z(\lambda)$ in eq.~\eqref{Zspec} translates into a very small change in $\Gamma_{CS}/(sT/N_c^2)$.

\begin{figure}[h!]
 \begin{center}
\includegraphics[scale=0.7]{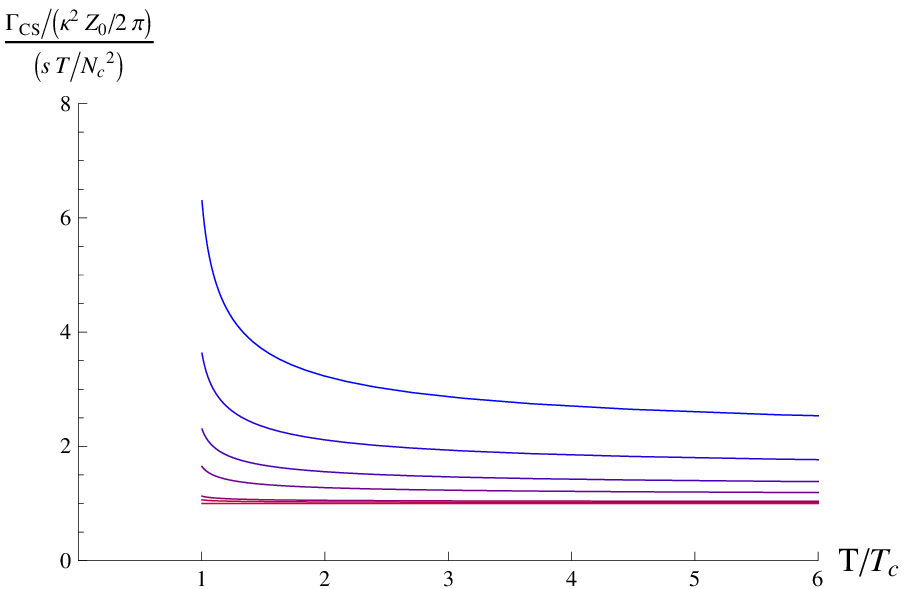}\hspace{1cm}
\includegraphics[scale=0.7]{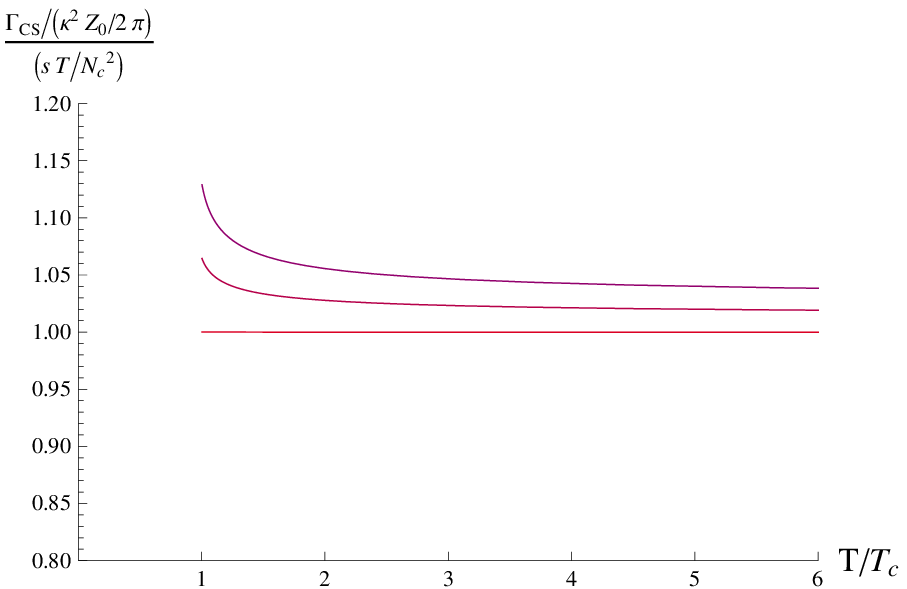}\\
(a) \hspace{5cm} (b)
 \end{center}
 \caption[]{(a) Our numerical results for $\Gamma_{CS}/(sT/N_c^2)$, normalized to the $T\to\infty$ value $\kappa^2 Z_0/2\pi$, as functions of $T/T_c$, for the $Z(\lambda)$ in eq.~\eqref{c1c4}, for different choices of the dimensionless parameters $(c_1,c_4)$. From the bottom (red) curve to the top (blue) curve, $(c_1,c_4) = (0,0.26), (0.5, 0.87), (1, 2.2), (5, 24), (10, 75), (20, 230), (40, 600)$. (b) Close-up of the curves for (from bottom to top) $(c_1,c_4)=(0,0.26), (0.5,0.87), (1,2.2)$. In all of these cases, as $T$ approaches $T_c$ from above $\Gamma_{CS}/(sT/N_c^2)$ increases by anywhere from $0.01\%$ up to a factor greater than six. The increase occurs mostly between $2T_c$ and $T_c$.}
\label{fig3b}
\end{figure}

\begin{figure}[h!]
 \begin{center}
\includegraphics[scale=0.8]{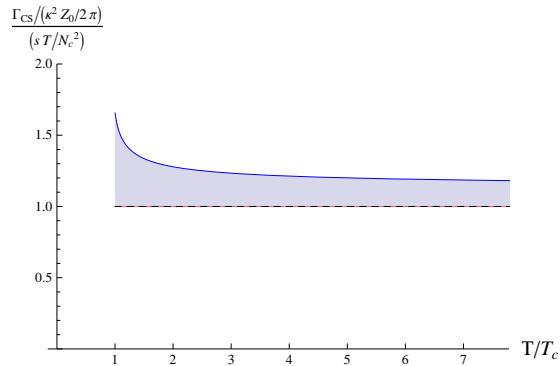}
 \end{center}
 \caption[]{Our numerical results for $\Gamma_{CS}/(sT/N_c^2)$, normalized to the $T\to\infty$ value $\kappa^2 Z_0/2\pi$, as functions of $T/T_c$, for the $Z(\lambda)$ in eq.~\eqref{c1c4} with $(c_1,c_4)$ constrained such that the holographic model reproduces the lattice results for axial glueball mass ratios to within one sigma: $0 \lesssim c_1 \lesssim 5$ and $0.06\lesssim c_4\lesssim 50$. A generic choice of $(c_1,c_4)$ within these limits will produce a curve inside the shaded region. The lower bound of the shaded region, given by the solid pink curve, has the lowest values, $(c_1,c_4) = (0,0.06)$, while the upper bound, given by the solid blue curve, has the largest values, $(c_1,c_4) = (5,50)$. At the upper bound we see that as $T$ approaches $T_c$ from above, $\Gamma_{CS}/(sT/N_c^2) \times (2\pi)/(\kappa^2 Z_0)$ increases by about $60\%$. The dashed line is the result for the optimal values $(c_1,c_4) = (0,0.26)$, as shown also in fig.~\ref{fig3}.}
\label{sigmafig}
\end{figure}

On the other hand, for the $Z(\lambda)$ in eq.~\eqref{c1c4}, for different values of the coefficients $c_1$ and $c_4$ we find more variation in $\Gamma_{CS}/(sT/N_c^2)$ as $T$ approaches $T_c$ from above, as shown in fig.~\ref{fig3b}. For example, if $c_1=40$ and $c_4=600$, then $\Gamma_{CS}/(sT/N_c^2)$ increases near $T_c$ by more than a factor of six. For all values of $c_1$ and $c_4$ that we considered, most of the increase occurs between $2T_c$ and $T_c$. Fig.~\ref{sigmafig} shows $\Gamma_{CS}/(sT/N_c^2)$, normalized to the $T\to\infty$ value $\kappa^2 Z_0/(2\pi)$, as a function of $T/T_c$ for values of $c_1$ and $c_4$ that reproduce the lattice results for axial glueball mass ratios to within one sigma, eq.~\eqref{onesigmalimits}. At the upper limits of the allowed $(c_1,c_4)$ values, namely $(c_1,c_4)=(5,50)$, we find that as $T$ approaches $T_c$ from above, $\Gamma_{CS}/(sT/N_c^2)$ increases by about $60\%$, with most of the increase occuring between $2T_c$ and $T_c$.

In heavy ion collisions at RHIC and LHC, $T$ reaches two to four times the QCD crossover temperature. We would thus like to know the value of $\gcs$ in QCD near the crossover temperature, which is a key ingredient determining the magnitude of any current produced via the CME~\cite{Kharzeev:2007jp}.\footnote{We thank D.~Kharzeev for a discussion on this point.} No controlled calculation of $\gcs$ from QCD at these temperatures exists, hence we turn to holography. Suppose we use $\N=4$ SYM as a holographic proxy for QCD near the crossover temperature. The result for $\gcs$ in large-$N_c$, strongly-coupled $\N=4$ SYM is~\cite{Son:2002sd},
\be
\gcs^{\N=4} = \frac{\lambda_t^2}{2^8 \pi^3} \,T^4.
\ee
Being a conformal field theory, $\N=4$ SYM has no phase transitions at non-zero $T$, so to obtain a sensible result we should consider the dimensionless quantity $\gcs/T^4$. As a crude estimate we take $\alpha_s\equiv g^2/(4\pi)=0.5$ and we use $N_c=3$, so that $\lambda_t = 6\pi$, in which case we find
\be
\label{CSconf}
\gcs^{\N=4}/T^4 \approx 0.045. \qquad (\lambda_t = 6\pi)
\ee
For a better estimate, let us consider $\gcs(T_c)/T_c^4$ in IHQCD. As discussed above, if $Z(\lambda)$ is monotonic in $\lambda$, then $\gcs/(s T/N_c^2)$ is bounded from below by its value in the $T\to\infty$ limit, eq.~\eqref{gcshighT}. We can obtain a lower bound on $\gcs(T_c)/T_c^4$ by using the large-$N_c$ YM lattice result for the entropy density at $T_c$~\cite{Lucini:2005vg}, $s(T_c) = 0.31 N_c^2 T_c^3$. Letting $\lambda_c$ denote the value of $\lambda_h$ at $T_c$, we find
\be
\label{CSTc}
\Gamma_{CS}(T_c)/T_c^4  = 0.31  \times \frac{\kappa^2 Z(\lambda_c)}{2\pi} >  0.31  \times \frac{\kappa^2 Z_0}{2\pi} = 1.64,
\ee
which is about 36 times larger than the $\N=4$ SYM estimate, eq.~\eqref{CSconf}. In fact, eq.~\eqref{CSTc} is closer to the perturbative QCD result, if we na\"ively extrapolate to $\alpha_s=0.5$: $\Gamma_{CS}(T)/T^4 \approx 30 \alpha_s^5 \approx 0.94$ (up to logarithms)~\cite{Arnold:1996dy,Bodeker:1998hm,Bodeker:1999gx,Moore:2010jd}. If we consider the $Z(\lambda)$ in eq.~\eqref{c1c4}, and constrain $c_1$ and $c_4$ to the values in eq.~\eqref{onesigmalimits}, then we can also place an upper bound on $\Gamma_{CS}(T_c)/T_c^4$, given by the solid blue curve in fig.~\ref{sigmafig}. For these choices of $Z(\lambda)$, we thus find
 \be
1.64\leq \Gamma_{CS}(T_c)/T_c^4 \leq 2.8.
\ee

Finally, we have also calculated $\Gamma_{CS}$ using the small black hole solutions~\cite{Gursoy:2008bu,Gursoy:2008za,Gursoy:2009jd}. Our results for those cases appear in the appendix. Although the small black hole branch is always thermodynamically disfavored, we can actually use the results for $\gcs/(sT/N_c^2)$ on the small black hole branch to argue quite generally that on the large black hole branch $\gcs/(sT/N_c^2)$ should increase as $T$ approaches $T_c$ from above. A similar argument also applies for the bulk viscosity, as discussed in ref.~\cite{Gursoy:2009kk}. The key result, shown in fig.~\ref{sBHfig} in the appendix, is that for $T>T_{min}$ $\gcs/(sT/N_c^2)$ is larger on the small black hole branch than on the large black hole branch, but the two branches meet at $T_{min}$. On the large black hole branch, then, $\gcs/(sT/N_c^2)$ must increase as $T\to T_{min}$ from above, in order to meet $\gcs/(sT/N_c^2)$ from the small black hole branch. In fact, we can show in full generality that on the large black hole branch $\gcs/(sT/N_c^2)$ must increase as $T\to T_{min}$: we simply take $(d/dT) (\Gamma_{CS}/(sT/N_c^2)) = (d\lambda_h/dT)(d/d\lambda_h) (\kappa^2Z(\lambda_h)/2\pi)$ and observe that by definition $(d\lambda_h/dT)$ diverges when $T\to T_{min}$, while $(d/d\lambda_h) (\k^2Z(\lambda_h)/2\pi)$ remains finite. Notice also that $\gcs/(sT/N_c^2)$ itself remains finite when $T\to T_{min}$. Given that $T_{min}$ is generally very close to $T_c$, we are then guaranteed that $\gcs/(sT/N_c^2)$ will be increasing as $T \to T_c$ from above, if we assume that $Z(\lambda)$ is monotonic as a function of $T$ between $T_{min}$ and $T_c$. In principle, $Z(\lambda)$ could exhibit maxima or minima for $T\in(T_{min},T_c)$, although such behavior seems un-natural. On the large black hole branch an increase of $\gcs/(sT/N_c^2)$ as $T \to T_c$ from above seems to be the generic behavior. We thus learn that the increase in $\gcs/(sT/N_c^2)$ in the vicinity of $T_c$ on the large black hole branch is tied to the existence of $T_{min}$, and hence to the existence of small black hole solutions. As argued in ref.~\cite{Gursoy:2008za}, the existence of small black hole solutions follows from the fact that the zero-temperature theory is confining. These arguments suggest that perhaps any confining, strongly-interacting, large-$N_c$ gauge theory with a (4+1)-dimensional holographic dual\footnote{Our arguments may not apply for (3+1)-dimensional confining theories obtained from higher-dimensional theories with compact spatial directions, such as the low-energy worldvolume theory on D4-branes with one spatial direction compactified and anti-periodic boundary conditions for fermions~\cite{Witten:1998zw}.} may exhibit an increase in $\Gamma_{CS}/(sT/N_c^2)$ in the vicinity of $T_c$.

\section{The Spectral Function}
\lab{CorrCS}
We now turn our attention to $G_R(\omega,\vec{k})$ with non-zero $\omega$ and $\vec{k}$. Generically $G_R(\omega,\vec{k})$ is a complex-valued function of the real variables $\omega$ and $\vec{k}$. A pole in $G_R(\omega,\vec{k})$ indicates a large response to an infinitesimal source for $q(x^{\mu})$, and is thus associated with a resonant excitation of the system. Being complex-valued, $G_R(\omega,\vec{k})$ is not directly observable. To study the excitations of our system, we thus turn to the spectral function, $-2 \,\textrm{Im} \,G_R(\omega,\vec{k})$, which is real and hence observable in principle.\footnote{Given $\textrm{Im} \, G_R(\omega,\vec{k})$ we can obtain $\textrm{Re} \, G_R(\omega,\vec{k})$ via a Kramers-Kroning relation, provided the large-$\omega$ and large-$|\vec{k}|$ asymptotics have been suitably regulated.} Typically, a pole in $G_R(\omega,\vec{k})$ produces a peak in the spectral function. In this section we initiate the study of these peaks in our system.

To be precise, we will compute $\textrm{Im} \, G_R(\omega,\vec{k})$. To do so, we will compute $G_R(\omega,\vec{k})$ using the membrane paradigm~\cite{Iqbal:2008by}, as explained in section~\ref{GammaCS}. In particular, we must solve eq.~\eqref{zetaeq}, which we reproduce here for convenience
\be
\label{zetaeq2}
\partial_r \zeta = \frac{\omega}{Z(\lambda(r)) b(r)^3 f(r)} \left[\zeta^2+ Z(\lambda(r))^2 b(r)^6 \left(1-f(r) \frac{\vec{k}^2}{\omega^2}\right) \right],
\ee
with the boundary condition in eq.~\eqref{bc},
\be
\label{bc2}
\zeta(r_h)= +i Z(\lambda_h) b(r_h)^3,
\ee
and then obtain $G_R(\omega,\vec{k})$ via eq.~\eqref{Kubohol},
\be
\lab{Kubohol2}
G_R(\o,\vec{k}) = -\kappa^2 M_p^3 \,\omega \lim_{r\to0} \zeta(r,\omega,\vec{k}).
\ee
We have not been able to solve eq.~\eqref{zetaeq2} exactly for all values of $\omega$ and $\vec{k}$, hence we turn to numerical solutions. In this section we exclusively use the $Z(\lambda)$ in eq.~\eqref{Zspec}, with $c_4=0.26$.

\begin{figure}[h]
 \begin{center}
\includegraphics[scale=0.5]{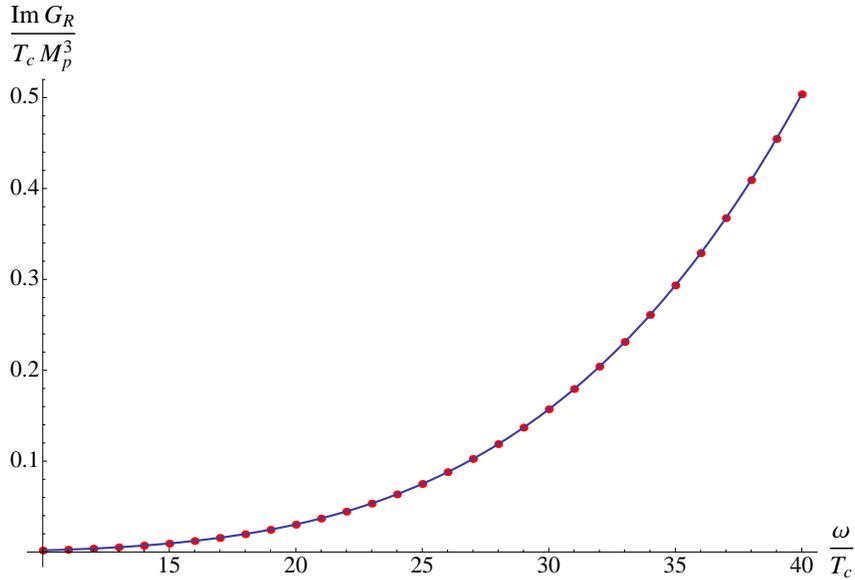}
\end{center}
 \caption[]{Our numerical results for $\textrm{Im}\,G_R(\omega,\vec{k}=0)/(T_c M_p^3)$ as a function of $\omega/T_c$, at $T_c$, for the $Z(\lambda)$ in eq.~\eqref{Zspec} with $\kappa^2 Z_0=33.25$ and $c_4=0.26$. The red dots are our numerical results while the solid blue curve is the function $(1.6 \times 10^{-7}) \times (\omega/T_c)^{4.051}$. Our results are clearly consistent with the expectation that $\textrm{Im}\,G_R(\omega,\vec{k}=0) \propto \omega^4$ at large $\omega$.}
\label{corr1}
\end{figure}

We consider first the case $\vec{k}=0$. Fig.~\ref{corr1} shows our numerical result for $\textrm{Im}\,G_R(\omega,\vec{k}=0)/(T_c M_p^3)$ at $T_c$ as a function of $\omega/T_c$. As we saw in section~\ref{GammaCS}, for $\omega$ sufficiently small, $\textrm{Im}\,G_R(\omega,\vec{k}=0) \propto \omega$. On the other hand, at asymptotically large $\omega$ we expect $\textrm{Im}\,G_R(\omega,\vec{k}=0) \propto \omega^4$ because in the UV the theory is conformally invariant and $q(x^{\mu})$ is dimension four. Our results are consistent with that expectation: fig.~\ref{corr1} shows that the function $(1.6 \times 10^{-7}) \times (\omega/T_c)^{4.051}$ provides an excellent fit to our data.

The $\omega^4$ scaling of $\textrm{Im}\,G_R(\omega,\vec{k}=0)$, and hence of $G_R(\omega,\vec{k}=0)$, at asymptotically large $\omega$ is a divergence in the coincidence limit of the two-point function that prevents the correlator from obeying the sum rules and dispersion relations typically used to give physical meaning to the poles of $G_R(\omega,\vec{k})$ in the complex $\omega$ plane, which require $G_R(\omega,\vec{k})$ to {\it vanish} at large frequency. Such a divergence may overwhelm peaks in $\textrm{Im}\,G_R(\omega,\vec{k})$, rendering them practically invisible.

\begin{figure}[h]
 \begin{center}
\includegraphics[scale=0.5]{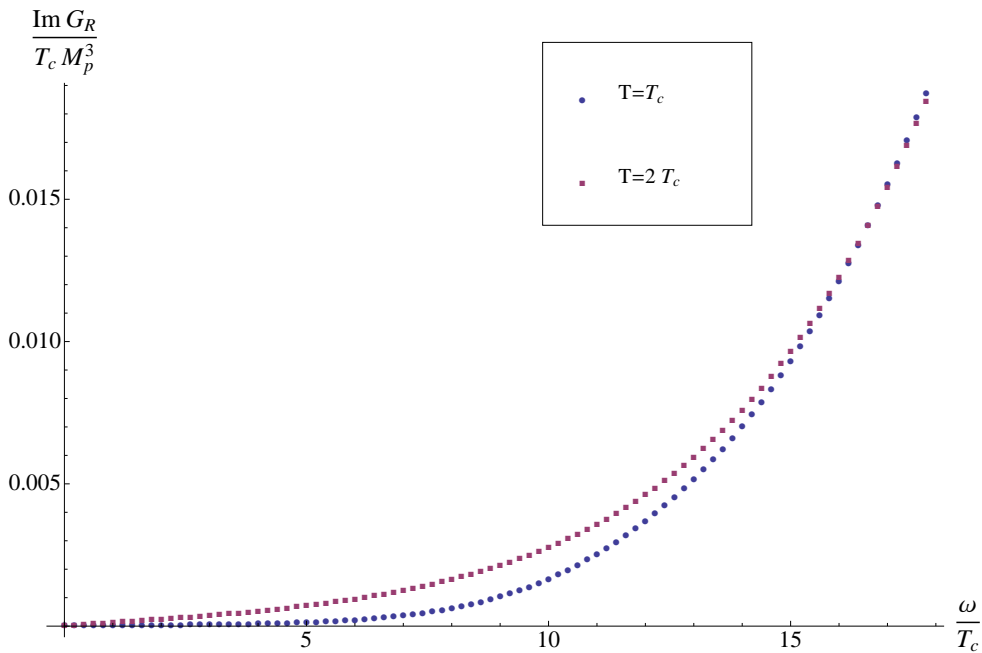}
\end{center}
 \caption[]{Our numerical results for $\textrm{Im}\,G_R(\omega,\vec{k}=0)/(T_c M_p^3)$ as a function of $\omega/T_c$, at $T_c$ (lower blue dots) and at $2T_c$ (upper red dots), for the $Z(\lambda)$ in eq.~\eqref{Zspec} with $\kappa^2 Z_0=33.25$ and $c_4=0.26$. A both $T_c$ and $2T_c$, for $\omega/T_c$ sufficiently large $\textrm{Im}\,G_R(\omega,\vec{k}=0) \propto \omega^4$.}
\label{corr2a}
\end{figure}

\begin{figure}[h]
 \begin{center}
\includegraphics[scale=0.5]{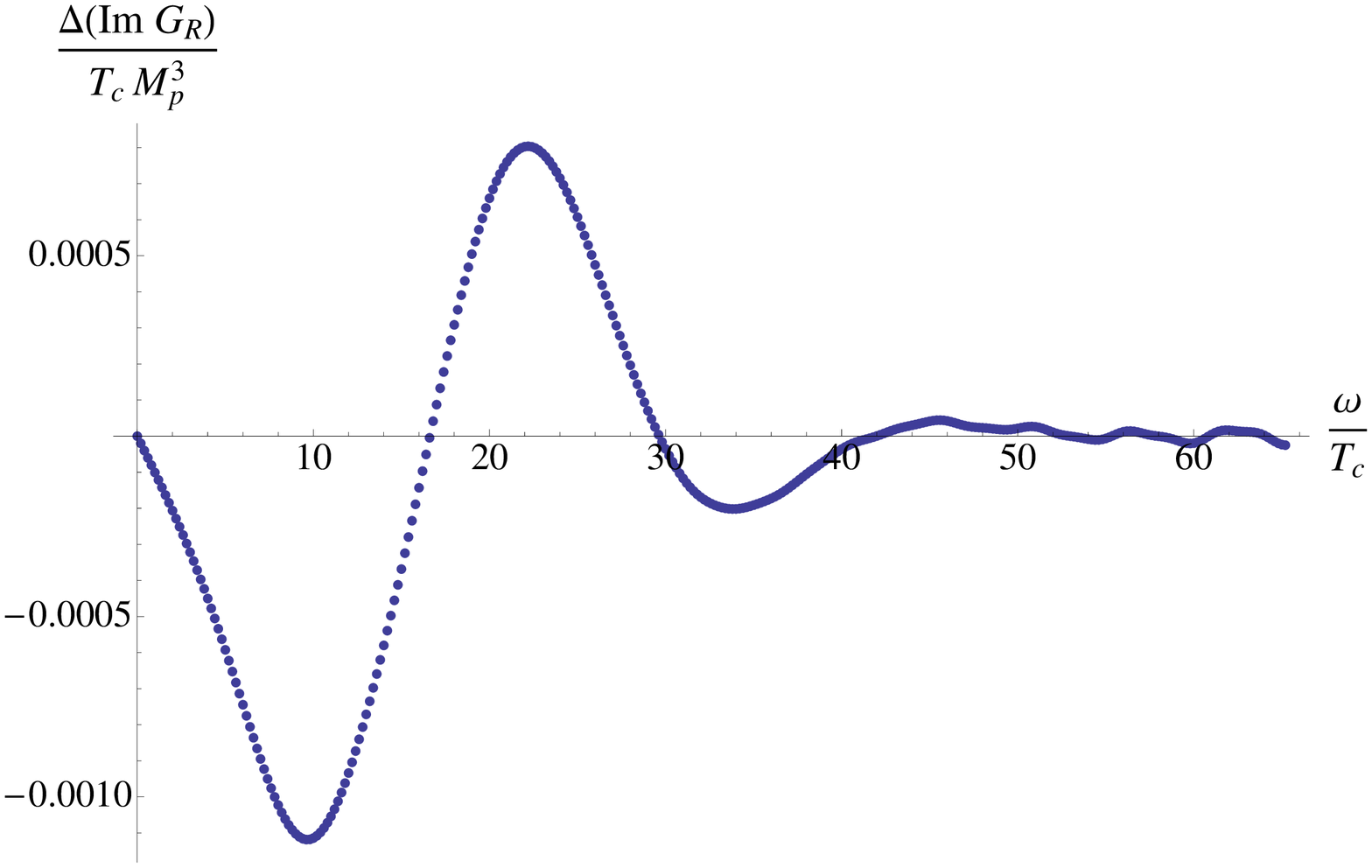}
\end{center}
 \caption[]{Our numerical results for the difference $\Delta \textrm{Im}\,G_R(\omega,\vec{k}=0;T_c,2T_c)/(T_c M_p^3)$ as a function of $\omega/T_c$, for the $Z(\lambda)$ in eq.~\eqref{Zspec} with $\kappa^2 Z_0=33.25$ and $c_4=0.26$. The difference goes to zero (within our numerical precision) as $\omega/T_c \to \infty$, as expected. The prominent minimum at $\omega/T_c\approx10$ and maximum at $\omega/T_c\approx 22$ indicate a shift in spectral weight with increasing $T$, presumably from the motion of a peak in the spectral function.}
\label{corr2b}
\end{figure}

One way to improve the large-$\omega$ behavior of $\textrm{Im}\,G_R(\omega,\vec{k})$ is to consider subtracted correlators. For example, one possible option is to determine the form of $\textrm{Im}\,G_R(\omega,\vec{k})$ at large $\omega$ exactly by solving eq.~\eqref{zetaeq2} in a WKB approximation, and then subtracting that large-$\omega$ form from all subsequent calculations of $\textrm{Im}\,G_R(\omega,\vec{k})$. That approach encounters ambiguities in sub-leading divergences in $\omega$, as discussed for example in ref.~\cite{Kajantie:2011nx}. We will instead eliminate the large-$\omega$ divergence by computing $G_R(\omega,\vec{k})$ at two different temperatures, $T_1$ and $T_2$, and then taking the difference,
\be
\lab{diffGR}
\Delta G_R(\omega,\vec{k}; T_1, T_2) \equiv \left.G_R(\omega,\vec{k})\right|_{T_2}  - \left.G_R(\omega,\vec{k})\right|_{T_1}.
\ee
We could also imagine subtracting the $T=0$ result for $G_R(\omega,\vec{k})$, that is, by taking $T_1=0$, but that is difficult to do numerically. When $T=0$, $G_R(\omega,\vec{k})$ is a sequence of delta-functions whose locations and amplitudes correspond to the masses and wave-function normalizations of axial glueballs.  We would need to subtract the enveloping function of this sequence of delta-functions, which is difficult to implement numerically. We will thus always consider $T_1,T_2\geq T_c$. Fig~\ref{corr2a} shows our numerical results for $\textrm{Im}\,G_R(\omega,\vec{k}=0)$ at two different temperatures, $T_c$ and $2T_c$, while fig.~\ref{corr2b} shows our numerical results for $\Delta\textrm{Im}\,G_R(\omega,\vec{k}=0;T_c,2T_c)$. In each figure we observe that the difference in $\textrm{Im}\,G_R(\omega,\vec{k}=0)$ between $T_c$ and $2T_c$ approaches zero as $\omega/T_c \to \infty$, at least within our numerical precision. Our numerical subtraction thus appears to be reliable, so we may interpret peaks in $\textrm{Im}\,G_R(\omega,\vec{k})$ as physical excitations.

From figs.~\ref{corr2a} and~\ref{corr2b}, we see that as $T$ increases from $T_c$ to $2T_c$, $\textrm{Im}G_R(\omega,\vec{k}=0)$ changes by at most $10\%$. Fig.~\ref{corr2b} also clearly reveals a minimum in $\Delta\textrm{Im}\,G_R(\omega,\vec{k}=0;T_c,2T_c)$ near $\omega/T_c\approx10$ and a maximum near $\omega/T_c \approx 22$, indicating a shift in spectral weight towards higher $\omega$ as $T$ increases. Indeed, fig.~\ref{corr2b} strongly suggests that a peak in the spectral function is moving to higher $\omega$ as $T$ increases. The location of the peak, at $\omega$ on the order of twenty times $T_c$, is roughly the same as the scale of the lightest $0^{-+}$ glueball mass at $T=0$, around $2600\,\textrm{MeV}$~\cite{Morningstar:1999rf}. In other words, fig.~\ref{corr2b} provides evidence that the plasma supports an excitation with roughly the same energy as the lightest $0^{-+}$ glueball at $T=0$. The width of the peak in fig.~\ref{corr2b} is about $10 T_c \approx 1300\,\textrm{MeV}$, so the excitation is reasonably long-lived.

\begin{figure}[h]
 \begin{center}
\includegraphics[scale=0.5]{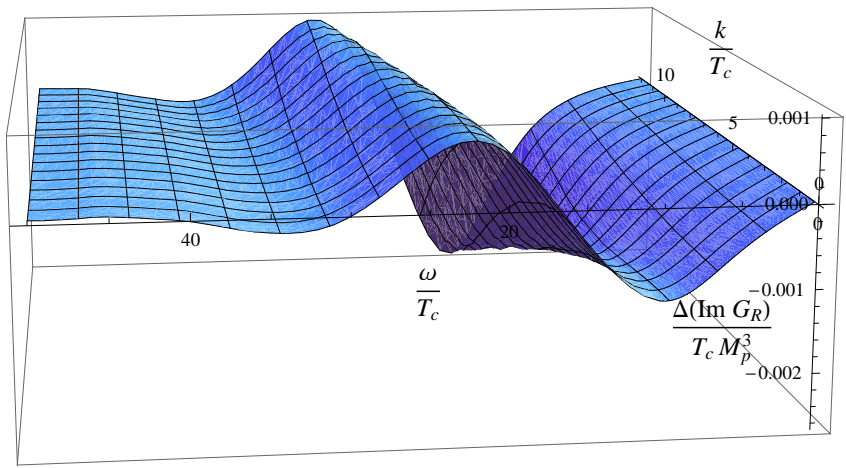}
\end{center}
 \caption[]{Our numerical results for $\Delta \textrm{Im}\,G_R(\omega,\vec{k}=0;T_c,2T_c)/(T_c M_p^3)$ as a function of $\omega/T_c$ and $|\vec{k}|/T_c$, for the $Z(\lambda)$ in eq.~\eqref{Zspec} with $\kappa^2 Z_0=33.25$ and $c_4=0.26$. As $|\vec{k}|$ increases up to $|\vec{k}|/T_c\approx 10$, the largest peak shifts from $\omega/T_c\approx 22$ up to $\omega/T_c\approx 30$. The width of the peak changes very little.}
\label{corr3}
\end{figure}

Figure \ref{corr3} shows our result for the subtracted correlator with non-zero $\omega$ and $|\vec{k}|$, using the same two temperatures as above. We observe that as $|\vec{k}|$ increases up to $|\vec{k}|/T_c\approx 10$, the largest peak shifts from $\omega/T_c\approx 22$ up to $\omega/T_c\approx 30$. Although this change in the position of the peak is roughly order one, the change in the shape of the peak is very mild. In particular, the width of the peak changes very little, indicating that the lifetime of the excitation stays nearly constant as $|\vec{k}|$ increases.

The typical time scale for dynamical processes in the QGP created in heavy ion collisions is about $1\,\textrm{fm/c} \approx (200\,\textrm{MeV})^{-1}$. Our results suggest the existence of a relatively long-lived excitation with energy on the order of $2600\,\textrm{MeV}$, corresponding to a time scale of about $0.1\,\textrm{fm/c}$. We cannot resist speculating that perhaps such an excitation, if present in the QGP, could dominate correlators of $q(x^{\mu})$ and hence many dynamical CP-odd phenomena. Regrettably, we will leave a detailed analysis of this excitation, and its effect on CP-odd physics, for the future.

\section{Discussion and Outlook} \label{discussion}

IHQCD is a state-of-the-art bottom-up holographic model for the low-energy physics of (3+1)-dimensional large-$N_c$ YM theory. In this paper we computed the retarded Green's function of the instanton density operator $q(x^{\mu})$ in the high-temperature, deconfined phase of IHQCD. Our primary motivation was to compute the Chern-Simons diffusion rate, $\gcs$, with the result in eq.~\eqref{gammaCS1}. In particular, our result for $\gcs$ is proportional to $Z(\lambda_h)$, where $Z(\lambda)$ is the normalization factor of the bulk axion action, and $\lambda_h$ is the value of the holographic 't Hooft coupling at the black hole horizon. A combination of available data for the topological susceptibility and axial glueball spectrum of large-$N_c$ YM, and glueball universality, are sufficient to determine the small and large $\lambda$ limits of $Z(\lambda)$~\cite{Gursoy:2007cb,Gursoy:2007er,Kiritsis:2009hu,Gursoy:2010fj}. We considered several forms for $Z(\lambda)$. Assuming that $Z(\lambda)$ is a monotonic function of $\lambda$, we found quite generally that $\gcs/(sT/N_c^2)$ is bounded from below by its value in the $T\to\infty$ limit and increases monotonically as $T\to T_c$ from above. Indeed, we presented an argument that the same will be true in many (3+1)-dimensional, confining, strongly-coupled, large-$N_c$ theories with holographic duals. For the $Z(\lambda)$ producing our optimal fit to the lattice results for the axial glueball spectrum, we found that the increase was only $0.01\%$. Fixing $Z(\lambda)$ completely by a least-squares fit to lattice results for the Euclidean two-point function of $q(x^{\mu})$, as explained in section~\ref{review}, is an important task for the future. We also presented evidence for a relatively long-lived excitation in the system with energy roughly on the order of the mass at $T=0$ of the lightest $0^{-+}$ glueball, which prompted us to speculate that perhaps such an excitation could dominate CP-odd phenomena in the QGP created in heavy ion collisions.

IHQCD is dual to pure large-$N_c$ YM, so an important goal for the future is to include the effects of quarks in the holographic calculation of $\gcs$. Some key questions are how the quark mass and chiral symmetry breaking affect $\gcs$. The axial and vector flavor $U(1)$ currents are dual to two $U(1)$ Maxwell fields in the bulk, and the quark mass operator is dual to a complex scalar field, a tachyon, that is bi-fundamental under these two gauge fields. In the bulk, the axion couples to the axial $U(1)$ gauge field and the to phase of the tachyon, as explained in refs.~\cite{Casero:2007ae,Jarvinen:2011qe}. A solution for the tachyon describing either a non-zero quark mass or chiral symmetry breaking can thus influence the axion and affect $\gcs$.

Introducing flavors fields would also enable us to compute holographically the current produced via the CME. A preliminary requirement is a bulk solution describing a magnetic field and a net chirality.

We plan to study these and other related issues in the future.

\acknowledgments{We would like to thank F.~Bruckmann, D.~Kharzeev, C.~Morningstar, H.~Panagopoulos, M.~Peardon, A.~Rago, A.~Sch\"{a}fer, D.T.~Son and L.~Yaffe for helpful conversations and correspondence. This work was supported in part by grants PERG07-GA-2010-268246, PIF-GA-2011-300984, the EU program ``Thales'' and "HERAKLEITOS II'' ESF/NSRF 2007-2013 and was also co-financed by the European Union (European Social Fund, ESF) and Greek national funds through the Operational Program ``Education and Lifelong Learning'' of the National Strategic Reference Framework (NSRF) under ``Funding of proposals that have received a positive evaluation in the 3rd and 4th Call of ERC Grant Schemes''. The research leading to these results has also received funding from the European Research Council under the European Community's Seventh Framework Programme (FP7/2007-2013) / ERC grant agreement no. 247252.}


\appendix
\renewcommand{\theequation}{\thesection.\arabic{equation}}
\addcontentsline{toc}{section}{Appendix: The Small Black Hole Branch}
\section*{Appendix: The Small Black Hole Branch}
\setcounter{equation}{0}
\renewcommand{\theequation}{\arabic{equation}}
\addcontentsline{toc}{section}{References}

As discussed in section~\ref{review}, when $T> T_{min}$ IHQCD admits two branches of black hole solutions, large black holes and small black holes~\cite{Gursoy:2008za}. In this appendix we compute $\gcs$ using the small black hole solutions.

\begin{figure}[h!]
 \begin{center}
  \subfigure[]{
        \includegraphics[width=0.43\textwidth]{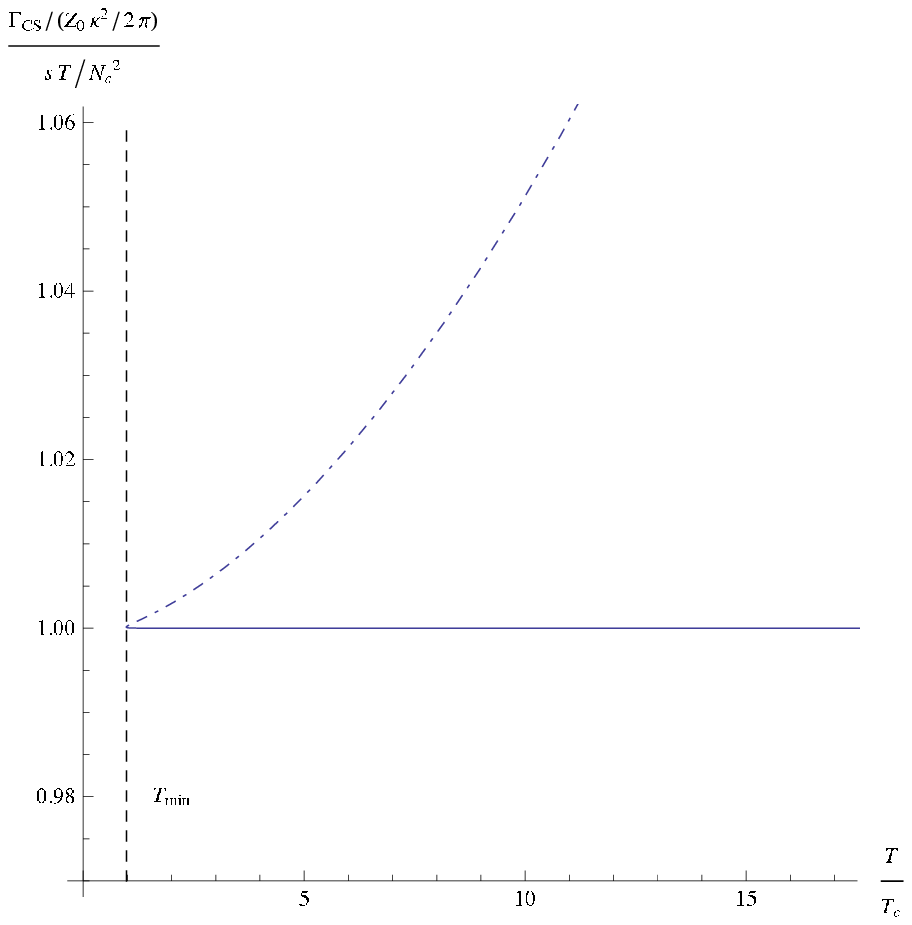}}\hspace{5mm}
   \subfigure[]{
        \includegraphics[width=0.43\textwidth]{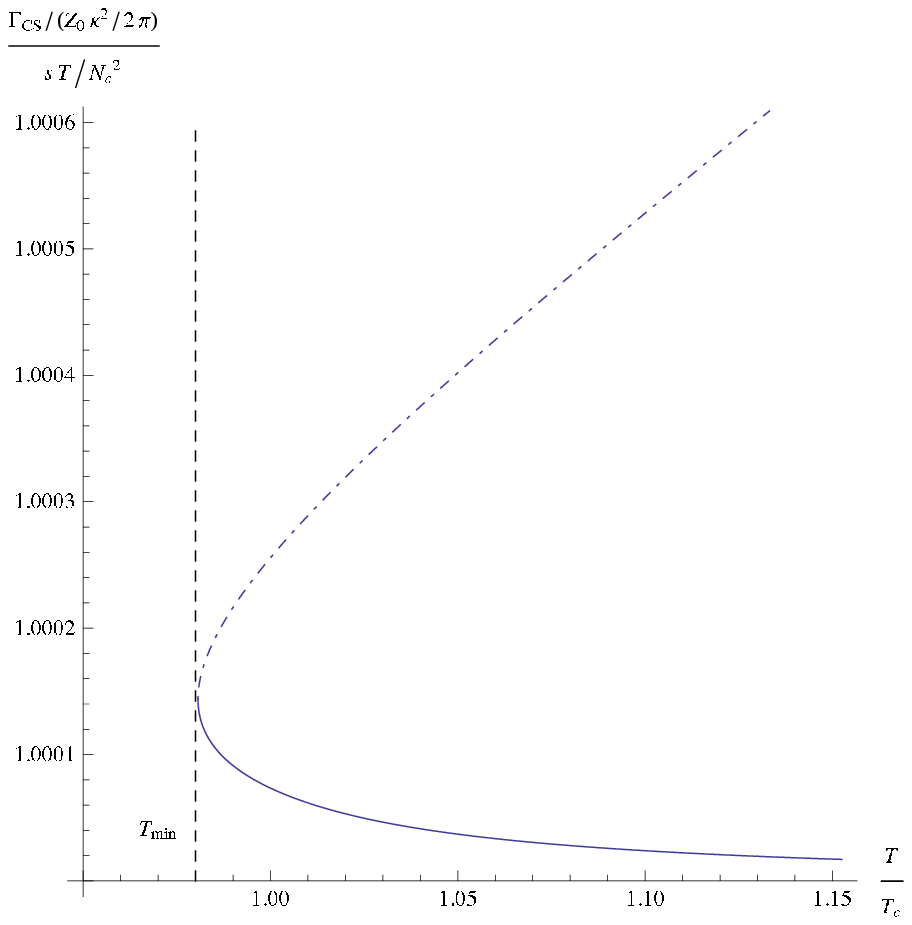}}
 \end{center}
 \caption[]{(a.) Our numerical result for $\Gamma_{CS}/(sT/N_c^2)$, divided by $Z_0 \kappa^2/(2\pi)$, as a function of $T/T_c$, for the $Z(\lambda)$ in eq.~\eqref{Zspec} with $c_4=0.26$. The upper dot-dashed blue curve is our result obtained from small black hole solutions, while the lower solid blue curve is the result obtained from large black hole solutions. Both curves begin at $T_{min}$, indicated by the vertical dashed black line, which is slightly below $T_c$. The result on the small black hole branch increases as $T$ increases, and in the $T\to\infty$ limit approaches the form in eq.~\eqref{eq:sbhgcs}. (b.) Close-up of (a.) near $T_{min}$.}
\label{sBHfig}
\end{figure}

We can determine the dependence of $\gcs$ on $T$ in the large-$T$ limit of the small black hole solutions as follows. For generality, we will consider a dilaton potential $V(\lambda)$ whose large-$\lambda$ asymptotic form is $V(\lambda) \propto \lambda^{4/3}\left(\log \lambda\right)^{P}$, with $P$ a non-negative real number. In the body of the paper we used $P=1/2$. From fig.~\ref{bhn}, we observe that for the small black hole solutions, when $T$ is large, $r_h$ is also large. When $r_h$ is large, $\lambda_h$ is also large, in which case we can approximate $Z(\lambda_h)\approx Z_0 c_4 \lambda_h^4$ and hence $\gcs / (sT/N_c^2)\approx \kappa^2 Z_0\,c_4\,\lambda_h^4/(2\pi)$. As shown in refs.~\cite{Gursoy:2007cb,Gursoy:2007er,Kiritsis:2009hu,Gursoy:2010fj}, in the $r\to\infty$ limit, $\lambda(r) \propto \exp(r^{1/(1-P)})\, r^{\frac{3}{4} \frac{P}{1-P}}$. Evaluating at $r_h$ gives us $\lambda_h$ in terms of $r_h$. From ref.~\cite{Gursoy:2008za} we know $r_h$ in terms of $T$ on the small black hole branch in the $r_h\to\infty$ limit, $r_h \propto T^{(1-P)/P}$. We thus find
\be
\label{eq:sbhgcs}
\frac{\gcs}{sT/N_c^2} \propto \frac{\kappa^2 Z_0 c_4}{2\pi} \, (T/T_c)^3 \, e^{C \left(T/T_c\right)^{\frac{1}{P}}},
\ee
where $C$ is a dimensionless positive constant that depends on the choice of $V(\lambda)$.

To compute $\gcs$ in the entire range $T_{min} < T < \infty$, we resorted to numerics. For the $Z(\lambda)$ in eq.~\eqref{Zspec} with $c_4=0.26$, our results appear in fig.~\ref{sBHfig}, where we see clearly that the result grows as $T$ increases, and in the $T\to\infty$ limit approaches the form in eq.~\eqref{eq:sbhgcs}. Fig.~\ref{sBHfig} also shows that the result for $\Gamma_{CS}/(sT/N_c^2)$ is always larger on the small black hole branch than on the large black hole branch, except at $T_{min}$ where the two are equal. This result is important for our argument at the end of section~\ref{GammaCS} that $\Gamma_{CS}/(sT/N_c^2)$ computed on the large black hole branch will increase as $T$ approaches $T_c$ from above.

\bibliography{biblio}
\bibliographystyle{JHEP}

\end{document}